\documentclass[twocolumn]{aastex63}

\usepackage{natbib}
\usepackage{amsmath}
\usepackage{graphicx}
\newcommand{\ddeg}{\hbox{$^{\circ}$}}

\newcommand{\ud}{2005~UD}

\newcommand{\farcsec}{\hbox{$.\!\!^{\prime\prime}$}}
\newcommand{\farcmin}{\hbox{$.\!\!^{\prime}$}}
\newcommand{\fdegree}{\hbox{$.\!\!^{\circ}$}}

\revised{\today}

\shorttitle{Spin Solution and Shape of 2005 UD}
\shortauthors{Kueny et al.}

\graphicspath{{./}{figures/}}

\begin{document}

\title{Implications for the Formation of (155140) \ud{} from a New Convex Shape Model}


\correspondingauthor{Jay K. Kueny}
\email{jkueny@arizona.edu}

\author[0000-0001-8531-038X]{Jay K. Kueny}
\affiliation{Lowell Observatory, 1400 W. Mars Hill Rd., Flagstaff, AZ 86001, USA}
\affiliation{Steward Observatory, University of Arizona, Tucson, 933 N. Cherry Ave., Tucson, AZ 85721, USA}
\affiliation{Wyant College of Optical Sciences, University of Arizona, 1630 E. University Blvd., Tucson, AZ 85721, USA}
\affiliation{National Science Foundation Graduate Research Fellow}

\author[0000-0001-7335-1715]{Colin Orion Chandler}
\affiliation{Department of Astronomy and Planetary Science, Northern Arizona University, PO Box 6010, Flagstaff, AZ 86011, USA}
\affiliation{National Science Foundation Graduate Research Fellow}
\affiliation{Dept. of Astronomy \& the DiRAC Institute, University of Washington, 3910 15th Ave. NE, Seattle, WA 98195, USA}

\author[0000-0002-6509-6360]{Maxime Devogèle}
\affiliation{Arecibo Observatory, University of Central Florida, HC-3 Box 53995, Arecibo, PR 00612, USA}

\author[0000-0001-6765-6336]{Nicholas Moskovitz}
\affiliation{Lowell Observatory, 1400 W Mars Hill Rd, Flagstaff, AZ 86001, USA}

\author[0000-0001-8434-9776]{Petr Pravec}
\affil{Astronomical Institute, Academy of Sciences of the Czech Republic, Fričova 1, CZ-25165 Ond\v{r}ejov, Czech Republic}

\author[0000-0002-1330-1318]{Hana Kučáková}
\affil{Astronomical Institute, Academy of Sciences of the Czech Republic, Fričova 1, CZ-25165 Ond\v{r}ejov, Czech Republic}
\affiliation{Institute of Astronomy, Faculty of Mathematics and Physics, Charles University, V Holešovickách 2, 18000 Prague, Czech Republic}
\affiliation{Research Centre for Theoretical Physics and Astrophysics, Institute of Physics, Silesian University in Opava, Bezručovo nám. 13, 74601 Opava, Czech Republic}

\author[0000-0002-0835-225X]{Kamil Hornoch}
\affil{Astronomical Institute, Academy of Sciences of the Czech Republic, Fričova 1, CZ-25165 Ond\v{r}ejov, Czech Republic}

\author[0000-0001-6098-6893]{Peter Kušnirák}
\affil{Astronomical Institute, Academy of Sciences of the Czech Republic, Fričova 1, CZ-25165 Ond\v{r}ejov, Czech Republic}

\author[0000-0002-5624-1888]{Mikael Granvik}
\affiliation{Department of Physics, P.O. Box 64, FI-00014 University of Helsinki, Finland}
\affiliation{Asteroid Engineering
Laboratory, Lule\aa{} University of Technology, Box 848, SE-98128 Kiruna, Sweden}

\author[0000-0002-4690-0157]{Christina Konstantopoulou}
\affiliation{Department of Astronomy, University of Geneva, Chemin Pegasi 51, 1290 Versoix, Switzerland}
\affiliation{Nordic Optical Telescope, Apartado 474, E-38700 Santa Cruz de La Palma, Spain}

\author[0000-0003-4670-9616]{Nicholas E. Jannsen}
\affiliation{Institute of Astronomy, Celestijnenlaan 200D bus 2401, B-3001 Leuven, Belgium}
\affiliation{Nordic Optical Telescope, Apartado 474, E-38700 Santa Cruz de La Palma, Spain}

\author[0000-0001-5221-0243]{Shane Moran}
\affiliation{Department of Physics and Astronomy, University of Turku, Vesilinnantie 5, FI-20500, Finland}
\affiliation{Nordic Optical Telescope, Apartado 474, E-38700 Santa Cruz de La Palma, Spain}

\author[0000-0002-6938-794X]{Lauri Siltala} 
\affiliation{Department of Physics, P.O. Box 64, 00014 University of Helsinki, Finland}

\author[0000-0002-8418-4809]{Grigori Fedorets} 
\affiliation{Department of Physics, P.O. Box 64, 00014 University of Helsinki, Finland}
\affiliation{Astrophysics Research Centre, School of Mathematics and Physics, Queen's University Belfast, Belfast BT7 1NN, UK}

\author[0000-0002-0535-652X]{Marin Ferrais}
\affil{Space sciences, Technologies \& Astrophysics Research (STAR) Institute University of Li\`{e}ge All\'{e}e du 6 Ao\^{u}t 19, B-4000 Li\`{e}ge, Belgium}
\affil{Aix Marseille Univ, CNRS, LAM, Laboratoire d’Astrophysique de Marseille, Marseille, France}

\author[0000-0001-8923-488X]{Emmanuel Jehin}
\affil{Space sciences, Technologies \& Astrophysics Research (STAR) Institute University of Li\`{e}ge All\'{e}e du 6 Ao\^{u}t 19, 4000 Li\`{e}ge, Belgium}

\author[0000-0003-1008-7499]{Theodore Kareta}
\affil{Lowell Observatory, 1400 W Mars Hill Rd, Flagstaff, AZ 86001, USA}

\author[0000-0002-2934-3723]{Josef Hanuš}
\affiliation{Institute of Astronomy, Faculty of Mathematics and Physics, Charles University, V Holešovickách 2, 18000 Prague, Czech Republic}

\begin{abstract}

(155140) \ud{} has a similar orbit to (3200) Phaethon, an active asteroid in a highly eccentric orbit thought to be the source of the Geminid meteor shower. Evidence points to a genetic relationship between these two objects, but we have yet to fully understand how 2005 UD and Phaethon could have separated into this associated pair. Presented herein are new observations of \ud{} from five observatories that were carried out during the 2018, 2019, and 2021 apparitions. We implemented light curve inversion using our new data, as well as dense and sparse archival data from epochs in 2005--2021 to better constrain the rotational period and derive a convex shape model of \ud{}. We discuss two equally well-fitting pole solutions ($\lambda = 116\fdegree6$, $\beta = -53\fdegree6$) and ($\lambda = 300\fdegree3$, $\beta = -55\fdegree4$), the former largely in agreement with previous thermophysical analyses and the latter interesting due to its proximity to Phaethon's pole orientation. We also present a refined sidereal period of $P_{\text{sid}} = 5.234246 \pm 0.000097$ hr. A search for surface color heterogeneity showed no significant rotational variation. An activity search using the deepest stacked image available of \ud{} near aphelion did not reveal a coma or tail but allowed modeling of an upper limit of 0.04 to 0.37~kg s$^{-1}$ for dust production. We then leveraged our spin solutions to help limit the range of formation scenarios and the link to Phaethon in the context of nongravitational forces and timescales associated with the physical evolution of the system.

\end{abstract}


\section{Introduction} 
\label{sec:intro}

Near-Earth asteroid (NEA) \ud{} is a kilometer-class object and is a potential flyby target of JAXA's DESTINY$^{+}$ mission,\footnote{\url{https://destiny.isas.jaxa.jp/science/}} scheduled to launch within the next decade. It was discovered in 2005 by the Catalina Sky Survey \citep{christensen_2005_2005} and was revealed to have an orbit similar to (3200) Phaethon and the Geminid meteor stream. A subsequent observational campaign revealed surface color variations as a function of rotational phase \citep{kinoshita_surface_2007}. Previous studies on the visible reflectance spectrum suggest that \ud{} is a B-type asteroid (\citealt{jewitt_physical_2006}; \citealt{devogele_new_2020}) though recent findings regarding the near-infrared spectrum by \cite{kareta_investigating_2021} and a phase curve analysis by \cite{huang_photometric_2021} contest this. It is in the Apollo dynamical class with a semimajor axis of 1.275 au, an eccentricity of 0.87, and an orbital inclination of $28\fdegree7$ (see Appendix \ref{appendixA} for a comprehensive reference table). Light curve inversion by \cite{huang_photometric_2021} using the Lommel--Seeliger ellipsoid method yielded a \ud{} spin pole solution of $(285\fdegree8\substack{+1.1 \\ -5.3}, -25\fdegree8\substack{+5.3 \\ -12.5})$ which is comparable to that of Phaethon (\citealt{hanus_3200_2018}; \citealt{kim_optical_2018}). A common origin with Phaethon continues to be extensively investigated (see, e.g., \citealt{devogele_new_2020}; \citealt{kareta_investigating_2021}; \citealt{maclennan_dynamical_2021}).

Asteroid (3200) Phaethon is a B-type NEA \citep{licandro_nature_2007} and exhibits short bursts of activity at perihelion suspected to be caused by thermal fracturing (\citealt{jewitt_activity_2010}; \citealt{jewitt_d_active_2012}). It has a semimajor axis of 1.271 au, an eccentricity of 0.89, and an orbital inclination of $22\fdegree3$. Phaethon is thought to be the parent body of the annual Geminid meteor shower \citep{whipple_1983_1983} although predicted upper limits to its current dust production rates cannot explain the inferred mass contained within the Geminid meteor stream (\citealt{ryabova_mass_2017}; \citealt{kasuga_wiseneowise_2022}).

Based on dynamical arguments (\citealt{ohtsuka_apollo_2006}; \citealt{hanus_near-earth_2016}; \citealt{maclennan_dynamical_2021}) \ud{} and Phaethon could have split from a larger precursor body at some point in the past which may explain the aforementioned mass discrepancy with the Geminids. A recent study of the Daytime Sextantids meteor shower (part of the Phaethon--Geminid stream complex) by \cite{kipreos_characterizing_2022} reinforces this by suggesting that this meteor stream, \ud{}, and Phaethon were created from a mutual breakup event. This common origin theory is further supported since B-type near-Earth objects are uncommon \citep{jewitt_physical_2006}, and the previously mentioned color variability discovered on \ud{} begs interesting implications for fresh surface material perhaps due to recent detachment. Analyses of \ud{} and Phaethon's polarimetric phase curve by \cite{devogele_new_2020} and \cite{ishiguro_polarimetric_2022} reveal similarities over broad phase-angle coverage, again hinting at a genetic relation between the pair. However, counterpoints to the common origin narrative were made by \cite{kareta_investigating_2021}, who found that these two objects have distinctly different spectral features in the near-infrared and by \cite{ryabova_asteroid_2019} based on dynamical tests probing the past 5000 yr.

In this work, we present further constraints on \ud{}'s rotation period, pole orientation, and shape model through light curve inversion using data from new 2018, 2019, and 2021 observations and archival data. We introduce the observations, data reduction, and shape modeling procedures in the next section. In Section \ref{sec:results}, we discuss the refined sidereal period and spin axis orientation for \ud{} as well as comment on the current state of a convex shape model. In Section \ref{sec:physical} we infer the most likely pole solution for \ud{} and present our search for surface color heterogeneity and activity. Section \ref{sec:discuss} leverages our spin solutions to inform possible formation scenarios for \ud{}. We then conclude with Section \ref{sec:summ} and encourage avenues for future work.

\section{Data Collection \& Processing}

We define ``light curve" as the time series of disk-integrated brightness of the asteroid collected at a single site in a single filter. We adopt the terms``dense" and ``sparse" to describe the two modalities of light curves used in our shape modeling process. Dense light curves feature photometric data points spaced closely in time relative to the rotational period of the object while sparse light curves typically contain interspersed points and light curve subsections fewer than about seven points per night, spanning greater than 30 days, and are usually the product of nightly astronomical surveys. In total, we used 79 dense and 5 sparse light curves of \ud{} from apparitions in 2005--2021 for our investigation. Of the set of dense light curves, we included 36 from \cite{devogele_new_2020}, 10 from \cite{warner_near-earth_2019}, 4 from \cite{jewitt_physical_2006}, 4 from \cite{kinoshita_surface_2007}, and the remaining from our own observations conducted in 2018, 2019, and 2021.

\subsection{Observations \& Photometry} 
\label{sec:obs}

We present photometric observations of \ud{} obtained using the following telescopes: the Ond\v{r}ejov Observatory 0.65 m telescope, the Danish 1.54 m Telescope,  North 0.6 m TRAnsiting Planets and PlanetesImals Small Telescopes (TRAPPIST-N), the 4.3 m Lowell Discovery Telescope (LDT), and the 2.6 m Nordic Optical Telescope (NOT). Our analysis also includes sparse data sets from various surveys (discussed below). Appendix \ref{appendixB} provides details about the observing circumstances.

We used the 4.3 m Lowell Discovery Telescope (LDT, located in Happy Jack, Arizona, USA) on the nights of UT 2019 October 19, 2019 November 18 and 2021 November 3. Images were captured using a broadband \textit{VR} filter (approximately encompassing the Johnson--Cousins $V$ and $R$ bands) and the Large Monolithic Imager, which features $6144 \times 6160$ pixels and a square $12\farcmin5$ field of view. This instrument samples at a pixel scale of $0\farcsec12$ pixel$^{-1}$ but was used in $3\times3$ binning mode. Exposure times ranged from 60 to 120 s for the first night, from 14 to 20 s for the second, and from 30 to 35 s for the last. Seeing conditions were very stable for the first and last night and stable at the 25\% level for the second night. We recorded median FWHM values of on-chip point sources of about $2\farcsec7$, $1\farcsec5$, and $2\farcsec3$ for the first, middle, and last night, respectively.

The 2.6 m NOT is located at the Spanish Observatorio del Roque de los Muchachos, La Palma, Canarias, Spain. For these data the NOT imaged with the Alhambra Faint Object Spectrograph and Camera (ALFOSC), which equips a nearly square $2048 \times 2064$ pixel detector sampled at $0.21"$ pixel$^{-1}$. Observations were carried out on the nights of UT 4 November 2019 and 18 November 2019 in the Sloan Digital Sky Survey (SDSS) $r$ filter and with a SDSS $g$-$r$-$i$ sequence, respectively. The images were subject to $2\times2$ binning and exposure times set at 30 s across both nights. Seeing varied typically from $1"$ to $2\farcsec5$ across both nights with conditions improving during the later half of the night for both runs.

Additional observations were collected by the robotic 0.6 m TRAPPIST-North \citep{jehin_trappist_2011} on the nights of UT 2019 November 24-26. TRAPPIST-N is located at the Oukaïmeden Observatory in the Atlas Mountains in Morocco. This telescope features an Andor iKON-L BEX2-DD CCD camera imaging through a Cousins $R$ filter. Additional instrument specifications include a $0\farcsec60$ pixel$^{-1}$ scale and a $22'$ square field of view. The images were binned $2\times2$ with exposure times set at 120 s. Seeing for the first night was variable with median on-chip values ranging from $\sim3"$ to $4\farcsec2$ with conditions improving on the second night, where values were in the range of $\sim2\farcsec7$ to 4". The final night unfortunately presented poor observing conditions, so we refrained from using data from this night in our analysis.

Observations in 2021 corresponding to nights UT October 27-30, as well as one night in 2018 on UT November 4, were carried out at La Silla Observatory using the Danish 1.54 m telescope (labeled "Danish" in Appendix \ref{appendixB}). The DFOSC instrument on this telescope has a deep depleted BI 2k$\times$2k sensor with 13.5 $\mu$m square pixels, and we used it unbinned. Integration times were between $60$ and $140$ s, and the telescope was tracked at half the apparent rate of the asteroid, providing star and asteroid source profiles in one frame.

For the remaining observations from the 2018 apparition we used the Ond\v{r}ejov Observatory 0.65 m telescope (labeled "Ond\v{r}ejov" in Appendix \ref{appendixB}) The 0.65 m is a reflecting telescope operated jointly by the Astronomical Institute of ASCR and the Astronomical Institute of the Charles University of Prague, Czech Republic. It uses a Moravian Instruments G2-3200 MkII CCD camera (with a Kodak KAF-3200ME sensor and standard \textit{BVRI} photometric filters) mounted at the prime focus. The CCD sensor has $2184\times1472$ square pixels (6.8 micron pitch) with microlenses and we imaged in $2\times2$ binning mode, providing $1\farcsec05$ pixel$^{-1}$ and a 19'$\times$12.8' field of view. Integration times were between 30 and 100 s, and we set the tracking at half-apparent rate of the asteroid.

We sourced the sparse data from the following:

\begin{enumerate}
  \item observations from the Catalina Sky Survey (CSS, \citealt{larson_css_2003});
  \item observations from the Pan-STARRS project \citep{chambers_pan-starrs1_2016};
  \item images from the ZTF project \citep{bellm_zwicky_2019} downloaded from the IRSA server (\url{https://irsa.ipac.caltech.edu/applications/ztf/}); and
  \item observations from the ATLAS project \citep{tonry_atlas_2018}.
\end{enumerate}

For all but the ZTF sparse data we utilized the calibrated chip-stage photometry (unpublished) and associated Julian dates reported on the Minor Planet Center (MPC)\footnote{\url{https://minorplanetcenter.net/db_search}}. Processing of the ZTF images is described in the following paragraph. 

We bias- and flat-field-corrected data from our new observations using standard techniques. We used the Python-based \textit{PhotometryPipeline} \citep{mommert_photometrypipeline_2017} to measure the photometry of these new data, as well as the ZTF sparse data. We note that neither the field stars nor \ud{} were trailed, and thus irregular photometry was not required for these new observations. A high-level overview of the pipeline workflow is as follows: astrometry using SCAMP \citep{bertin_automatic_2006}, which utilizes the VizieR catalog service \citep{ochsenbein_vizier_2000} to perform image registration via the Gaia Data Release 2 catalog \citep{gaia_collaboration_gaia_2018}; point-source extraction using Source Extractor \citep{bertin_sextractor_1996}; photometric zero-point calibration using the Pan-STARRS DR1 catalog \citep{flewelling_pan-starrs1_2020}; and distilling the calibrated photometry by using the object's position in the frame as returned through a query to the JPL Horizons system \citep{giorgini_jpls_1996}. Additionally, the pipeline executed photometric calibration using stars with solar-like colors in the same frame from the SDSS DR9 catalog \citep{ahn_ninth_2012}, where the color thresholds were set at $(g - r) = 0.44 \pm 0.2$ and $(r - i) = 0.11 \pm 0.2$. Using the curve-of-growth procedure outlined in \cite{mommert_photometrypipeline_2017}, the pipeline determined a best-fit aperture radius of 2.84--4.53 binned pixels ($1\farcsec22$ to $1\farcsec94$) for the NOT data and 3.26--5.37 binned pixels ($1\farcsec17$ to $1\farcsec93$) for the LDT data. The TRAPPIST data were the only exception to this where we manually set apertures of 5-pixel radius (3").

Data from the 2018 and 2021 apparition that were taken with the Ond\v{r}ejov Observatory 0.65 m and Danish 1.54 m telescopes were subject to a custom aperture photometry program Aphot + Redlink developed by Petr Pravec and Miroslav Velen. In short, the software performs a semiautomated routine to select optimal apertures for the photometry. Star-like sources in the science frames are calibrated in the Johnson--Cousins \textit{V-R} system with standard stars from \cite{landolt_ubvri_1992} facilitating 0.01 mag precision in photometric conditions.

\begin{figure*}
	\centering
	\includegraphics[width=1.0\linewidth]{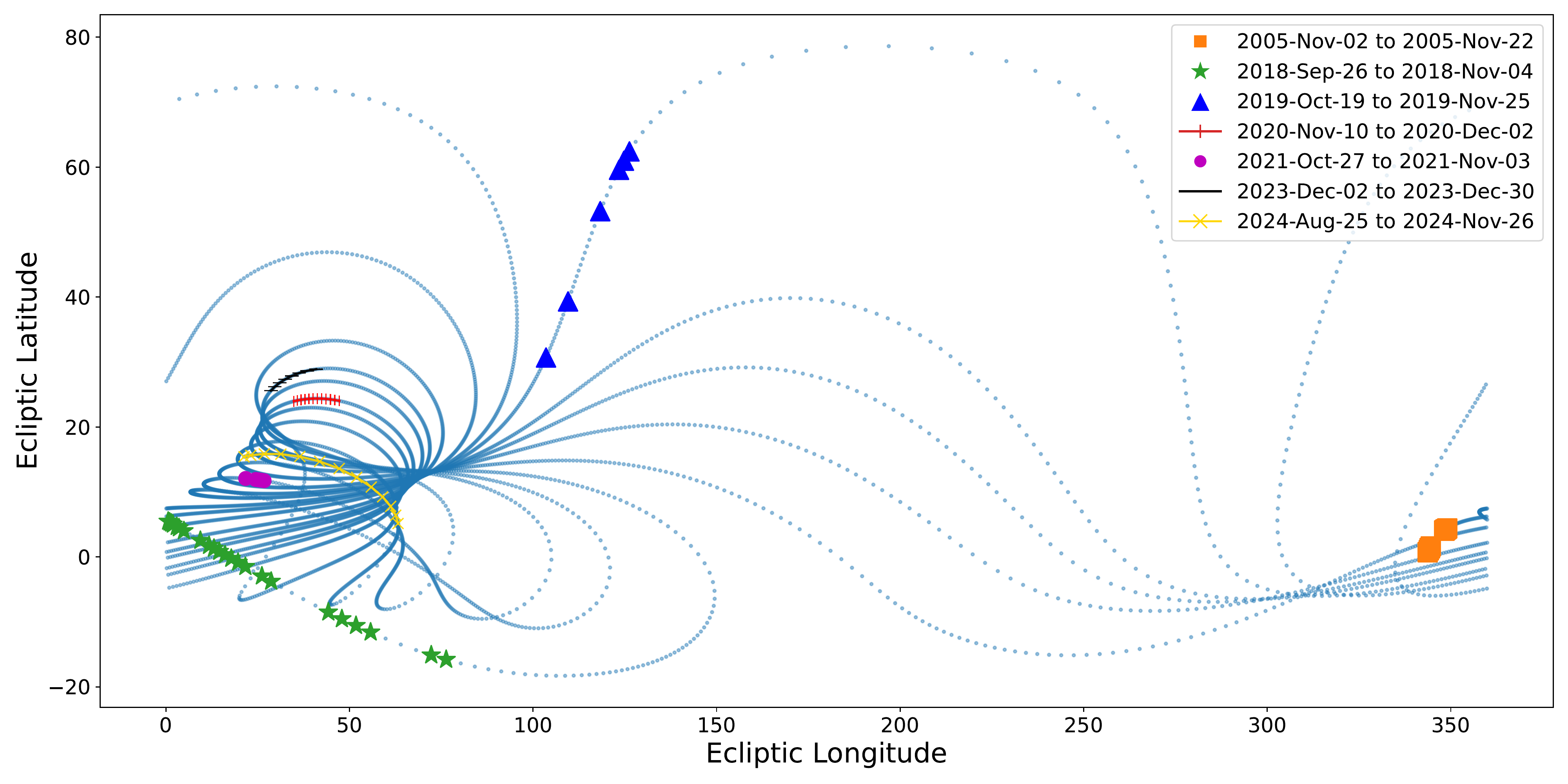}
	\caption{Geocentric ecliptic coordinates at the time observations of \ud{} were obtained. The small blue circles indicate the ecliptic coordinate of \ud{} from 2005 September 5 to 2024 December 31 with a resolution of 1 day. The 2005, 2018, 2019, and 2021 dense light curves included in our analysis are shown as the orange squares, green stars, blue triangles, and magenta circles, respectively. A past viewing opportunity in 2020 November is represented as the red vertical hash region. Additional viewing opportunities of \ud{} in late 2023 and late 2024 are represented as the black horizontal ticks and yellow cross regions, respectively. Ecliptic coordinates of \ud{} were obtained from JPL Horizons \citep{giorgini_jpls_1996}.}
	\label{fig:obs}
\end{figure*}

Appendix \ref{appendixB} includes further details of all light curves used in our analysis. Figure \ref{fig:obs} shows ecliptic coordinates corresponding to our new observations, as well as future apparitions. 

\section{Light curve Inversion} 
\label{sec:results}

Light curve inversion has been used to ascertain spin states of NEAs; see models for, e.g., Phaethon (\citealt{hanus_near-earth_2016}; \citealt{hanus_3200_2018}), Cuyo \citep{rozek_physical_2019}, and Apollo (\citealt{kaasalainen_acceleration_2007}; \citealt{durech_new_2008}). In addition to the necessity of good quality data (i.e., high signal-to-noise ratio), a unique solution requires data obtained across a broad range of viewing geometries (i.e., sampling reflectance data from as much of the surface of the object as possible). To determine the shape (expressed as a convex polyhedron) and pole solution of \ud{} as well as refine its sidereal rotation period, we used the \textit{convexinv} program, described in \cite{kaasalainen_optimization_2001-1} and \cite{kaasalainen_optimization_2001}.

Prior to carrying out light curve inversion, we formatted the data (see the \textit{convexinv} documentation) into standard ``blocks". We employed the \textit{astropy}-affiliated \citep{tardioli_constraints_2017} \textit{astroquery} tool to obtain Sun and Earth $xyz$ vector components, centered on \ud{} and expressed in au, for each data point via the JPL Horizons service \citep{giorgini_jpls_1996}. Additionally, all light curves were normalized to unity and converted to flux units before correcting observation times for light-travel time. As a final step, all flux values were range-corrected to 1 au from the Earth and Sun.
  
Of the original light curve set from \cite{devogele_new_2020}, we discarded five light curves obtained at the Lowell Observatory 0.79 m National Undergraduate Research Observatory telescope (31in henceforth) and one light curve from TRAPPIST-N due to high photometric noise or having temporally overlapping data from a superior instrument (although our model light curves were able to reproduce these data). These six light curves correspond to observations on UT 2018 September 27, October 6, October 10, October 15, October 16, and October 17. Additionally, we exclude the first half of one LDT light curve from our new data taken on UT 2019 October 18 from the shape modeling procedure because light curve predictions from our best-fit shape models did not agree with the uncharacteristic 0.5 mag amplitude; the cause for the discrepancy in this light curve with our models is unknown. Preparing the sparse data sets included rejecting any data points taken near the magnitude limit of the specific instrument and plotting the observed intensities vs. the associated phase angles and performing a sigma-clipping routine to eliminate outliers.

\subsection{Rotational Period}
\label{sec:period}


To get a reliable shape solution, it is imperative to constrain the rotational period of the object. The parameter space is riddled with local minima due to the rotational period, which, since \textit{convexinv} is a gradient-based algorithm, will cause the optimizer to get trapped and converge to an ill-fitting solution. The spacing of these local minima $\Delta P$ (in hours) in the period-space $\chi^2$ spectrum is given roughly by $\Delta P \approx P^2/(2T)$, where $P$ is the rotational period of the object and $T$ is the time span of the entire data set \citep{kaasalainen_optimization_2001}. We ensured that the step size of the period search did not increase significantly above this value to prevent missing the correct period. We define a unique period or spin solution if its $\chi^2$ (see Section 3.1 in \citealt{kaasalainen_optimization_2001-1} for how this test statistic is calculated) is lower than that of all other candidate solutions by at least 10\% (also used in \citealt{hanus_study_2011}). Note that while the criterion employed in \cite{hanus_3200_2018} suggests that a $\chi^2$ threshold of $\sim5\%$ above the minimum is valid for our dataset, we opt for the more rigorous 10\% threshold to make the global minimum more distinct.

Prior to running \textit{period\_scan} we specified a narrow period search window centered on the literature value and assigned weights to each light curve to improve the goodness of fit; noisy and sparse light curves were assigned lower weights. We optimized the individual weights $W_{\text{LC}}$ for each remaining dense light curve quantitatively using $W_{\text{LC}} = 1/{\text{rms}^2}$, where the rms error is obtained by fitting a Fourier series to each light curve. Next, all nonzero weights for dense light curves were then multiplied by a scale factor to bring the sum of the dense light curve weights equal to the number of nonrejected dense light curves (in our case 78). For the 2005, 2019, and 2021 apparitions we fit a Fourier series up to the seventh order to each light curve. For the 2018 apparition, we limited the fitted Fourier series to second order as a form of regularization since these 2018 data compose $\sim80\%$ of our dense light curves. Performing various shape modeling trials showed that fully optimizing the weights for all dense light curves biased the model heavily to the 2018 apparition by suppressing the weights of the light curves from the other apparitions. This is likely because data from the other apparitions, particularly from 2005 and 2021, are quite limited in both quantity and quality. Further, because of \ud{}'s relatively smooth and invariant sinusoidal light curve (see Figure \ref{fig:plots2019} for an example), we determined that a second order Fourier series was sufficient in penalizing lower-quality light curves without significantly underestimating the weights of the high-quality ones. For the sparse light curves, we assign weights at the 10\% or 20\% level (see Section \ref{sec:pole}) by taking the average of the lowest 10\% or 20\% of weights of the dense light curves.

\begin{table}
\centering
\caption{Comparison of Previous Rotational Period Estimates for \ud{} with This Work.}
\begin{tabular}{ccc}
\hline \hline Period   & Uncertainty                       & Source \\
(hr)     & (hr)                               &        \\ \hline
5.23     &                                   & \cite{jewitt_physical_2006}     \\
5.2492   &                                   & \cite{kinoshita_surface_2007}      \\
5.231    & $\pm0.034$                        & \cite{sonka_155140_2019}      \\
5.235    & $\pm0.005$                        & \cite{devogele_new_2020}      \\
5.2340   & $\substack{+0.00004 \\ -0.00001}$ & \cite{huang_photometric_2021}      \\
5.234246 & $\pm0.000097$                     & This work      \\ \hline
\end{tabular}
\label{tab:periods}
\end{table}

\begin{figure}
	\centering
	\includegraphics[width=1.0\linewidth]{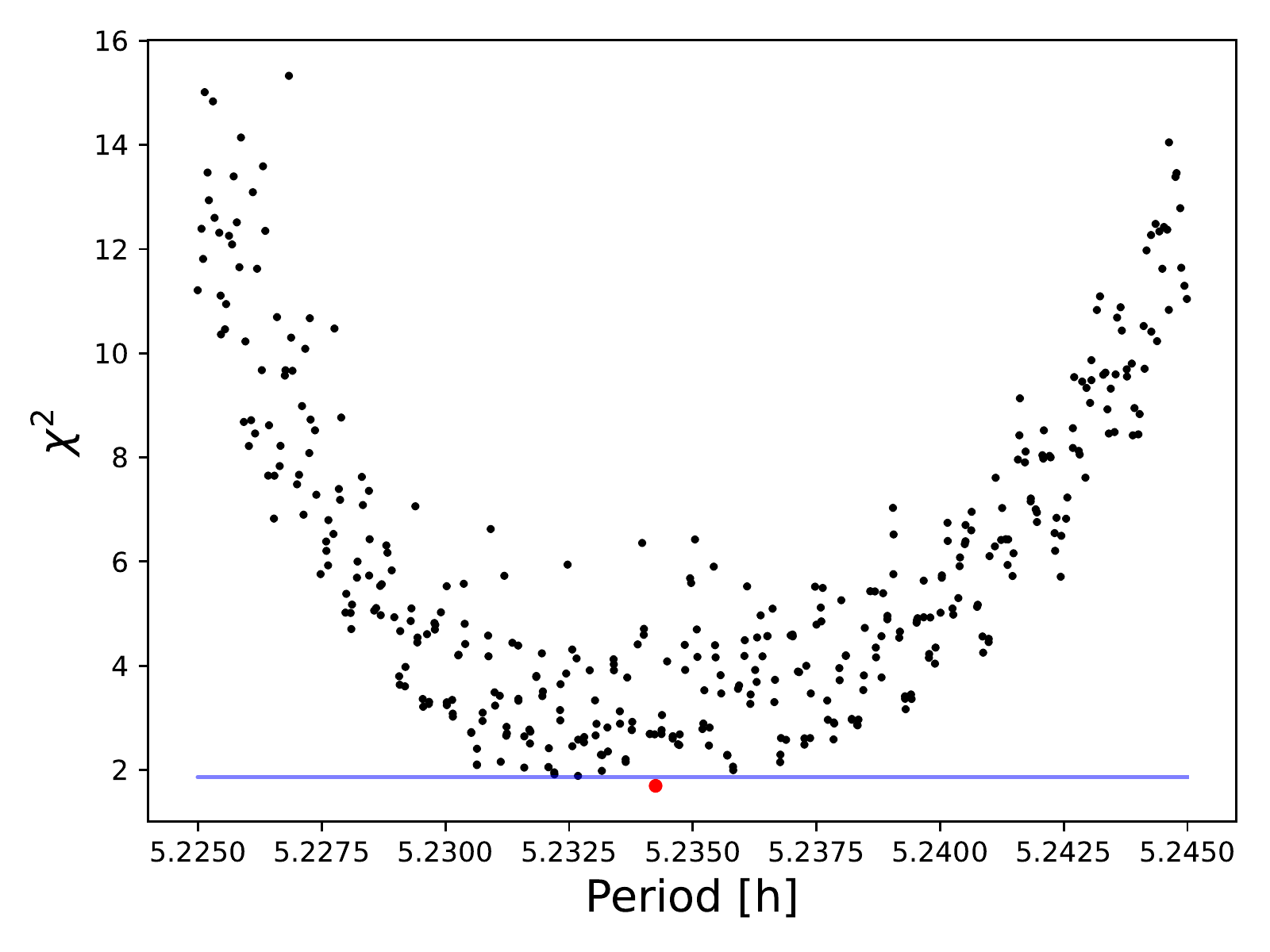}
	\caption{The \textit{period\_scan} results from our set of 84 light curves with optimized weights. The black points represent local minima in the sidereal period, pole orientation, and shape parameter space. The blue horizontal line shows the $\chi^2$ value 10\% higher than the global minimum to which we consider points underneath to be viable solutions. The only solution that satisfies this criterion for the rotational sidereal period with our dataset is $P_{\text{sid}} = 5.234246 \pm 0.000097$ and is shown in red.}
	\label{fig:period}
\end{figure}

The \textit{period\_scan} results within a search range of 5.225 and 5.245 hr produced a unique sidereal rotational period value of $P_{\text{sid}} = 5.234246 \pm 0.000097$ hr (Figure \ref{fig:period}). A comparison of our period solution with those from previous studies is presented in Table \ref{tab:periods}. A previous period search for \ud{} conducted in \cite{devogele_new_2020} suggested the possibility of a three-peaked light curve corresponding to a rotation period around 7.85 hr. With the addition of our new data, we note significant reduction in the goodness of fit for periods in this vicinity. As such, we can now exclude sidereal period solutions in this range and surmise that this is an alias.

\subsection{Spin and shape solutions}
\label{sec:pole}

Using our period solution, we followed the standard \textit{convexinv} recipe to derive a shape model. We specified 48 initial pole orientations isotropically distributed on the celestial sphere at $30\ddeg$ spacing, which is later optimized during the inversion procedure. The uncertainties for the sparse data from the MPC (all but the ZTF data) are not reported. Thus, the contribution of each sparse dataset toward convergence was assessed by performing a coarse grid search across all sparse data combinations, where each individual sparse light curve was assigned a 0\%, 10\%, or 20\% weight. By scanning through these weights, we aimed to minimize the relative $\chi^2$ value for the dense light curves only. Following this test,  we assigned a weight of 10\% to all but the ZTF data (assigned 20\%) as this specific combination produced the lowest $\chi^2$. This combination is not surprising since we reduced the ZTF sparse data ourselves and thus had the knowledge to reject individual data points based on error, source blending, etc. The ZTF data also had the largest phase-angle coverage (i.e., more viewing geometries) of the sparse data sets, covering $\sim100\ddeg$ in phase over a 3 yr span.

\begin{table}
    \raggedright
    \caption{Candidate Spin Solutions after Performing Light Curve Inversion.}
    \begin{tabular}{lcccc}
      \hline \hline  Solutions & $\lambda$ & $\beta$ & $\chi^2$ & $\mathrm{rms}$ \\
            & (deg)    &   (deg)  &      &     \\ \hline 
        1*                            & 144.0                 & -31.8        & 3.0542         & 0.0202    \\
        2                            & 116.6                & -53.6         & 3.1037           & 0.0203      \\
        3                            & 300.3                 & -55.4         & 3.1190           & 0.0204      \\
        4*                            & 62.5                & -44.2        & 3.2048             & 0.0207         \\
        \hline
    \end{tabular}
    
    \footnotesize
    \vspace{2mm}
    \textbf{Note.} $^*$ Rejected due to nonphysical shape. $\chi^2$ and $\mathrm{rms}$ quantify fits to dense light curves only.
    \label{tab:solutions}
\end{table}

Following this, we were left with four probable spin solutions, which are listed in Table \ref{tab:solutions}. All four shape models had a small dark facet area (below 1\% of the total facet area), which is needed to preserve convexity, so we were not able to eliminate any of these candidate solutions using this metric. Further, this suggests that nonconvex features (e.g., concavities) do not occupy a significant area on \ud{}'s surface. Following this, we computed the inertia tensors (see \citealt{dobrovolskis_inertia_1996} for a full description) of the shape models to discover that Solution 1 and Solution 4 had an inertial axis significantly misaligned from the z-axis, so we rejected these solutions on the basis of being nonphysical (i.e. there is no evidence that \ud{} is in nonprincipal axis rotation). However, one caveat is that if \ud{} is not uniform in color (see Section \ref{sec:heterogeneity}) or density, then our convex approximation may contain systematic errors which could manifest as the aforementioned inertial axis misalignment.

This left us with Solution 2, $(\lambda_2, \beta_2) = (116\fdegree6 \pm 2\fdegree2, -53\fdegree6 \pm 4\fdegree7$), and Solution 3, ($\lambda_3, \beta_3) = (300\fdegree3 \pm 2\fdegree5 -55\fdegree4 \pm 2\fdegree2$) ($1\sigma$ errors), which are equally well fitting. Errors were estimated by agitating various parameters during dozens of trial runs and investigating the effects on $\chi^2$ values. Solution 3 is interesting because constraints on the pole orientation of (3200) Phaethon place it in the range of $308\ddeg \lesssim \lambda \lesssim 322\ddeg$ and $-40\ddeg \lesssim \beta \lesssim -52\ddeg$ (\citealt{kim_optical_2018} and \citealt{hanus_3200_2018}), which is within our $3\sigma$ error. Further discussion on the implications of these pole solutions for the formation of \ud{} is given in Section \ref{sec:discuss}. 

Recent shape modeling efforts using the Lommel--Seeliger ellipsoid method by \cite{huang_photometric_2021} yielded two candidate pole solutions for \ud{}: (1) $(72\fdegree6\substack{+4.2 \\ -7.3}, -84\fdegree6\substack{+6.2 \\ -2.1})$ and (2) $(285\fdegree8\substack{+1.1 \\ -5.3}, -25\fdegree8\substack{+5.3 \\ -12.5})$. Pole 2 is favored as their preferred solution and is both comparable to Phaethon's and largely in agreement with our Solution 3 with overlap in longitude within $3\sigma$ errors.

\begin{figure}
    \centering
    \includegraphics[width=\linewidth]{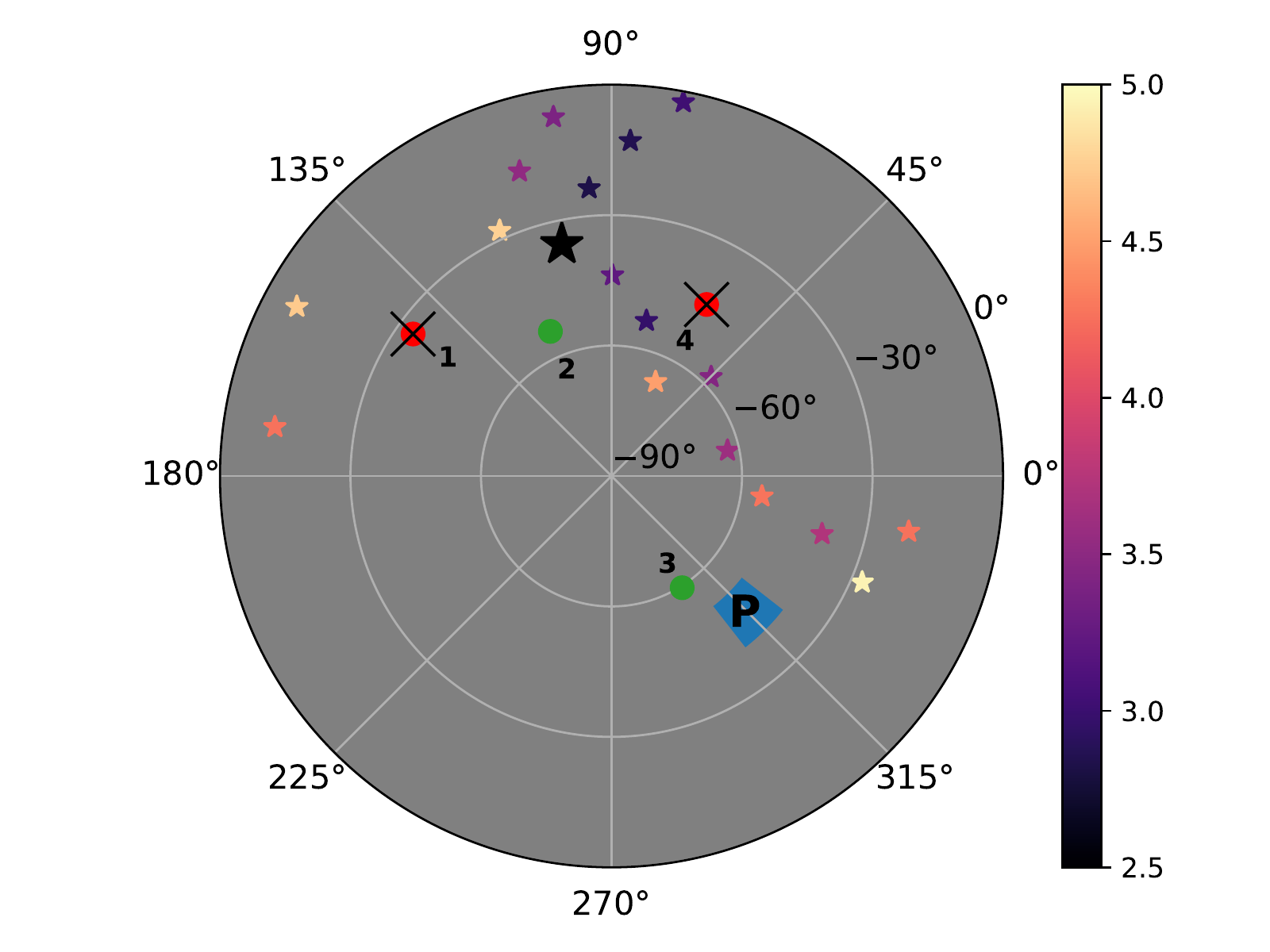}
    \caption{Polar projection map comparing our four pole solutions from light curve inversion, candidate thermophysical pole solutions from \cite{devogele_new_2020}, and Phaethon's spin pole region from \cite{hanus_3200_2018} and \cite{kim_optical_2018}. The map is centered on the south ecliptic pole. The green circless show accepted light curve inversion solutions, while the crossed-out red circles represent our rejected solutions. The color bar represents $\chi^2$ values of the thermophysical solutions, which are represented as stars; only retrograde solutions with $\chi^2 \leq 5$ are shown for clarity. The best-fitting thermophysical solution is represented by the large black star. Phaethon's spin axis orientation with approximate uncertainty is marked as the letter $P$ surrounded by the blue extended region.}
    \label{fig:solutions}
\end{figure}

Thermal data of \ud{} from two different epochs were obtained during the Near-Earth Object Wide Infrared Explorer (NEOWISE) reactivation mission \citep{mainzer_initial_2014} using the two shortwave filters (W1: $3.1 \mu$m; W2: $4.6 \mu$m). Thermophysical modeling of these data was presented in \cite{devogele_new_2020}. The best-fit solution from the thermophysical model ($\lambda_{\mathrm{TPM}} = 102\ddeg \pm 20\ddeg$ and $\beta_{\mathrm{TPM}} = -35\ddeg \pm 30\ddeg$) is consistent with Solution 2. Given that the other pole solution candidates (Solutions 1, 3, and 4)  were inconsistent with the predicted longitude from the thermophysical analysis and Solution 2 consistently attained the lower $\chi^2$ among numerous trial runs, we adopt $\lambda_p = 116\fdegree6 \pm 2\fdegree3$, $\beta_p = -53\fdegree6 \pm 5\fdegree4$ as our preferred solution. A comparison between our pole solutions from light curve inversion, solutions from the above thermophysical modeling, and Phaethon's pole is illustrated in Figure \ref{fig:solutions}.

\begin{figure*}
\centering
\begin{tabular}{cc}
        \includegraphics[width=0.475\linewidth]{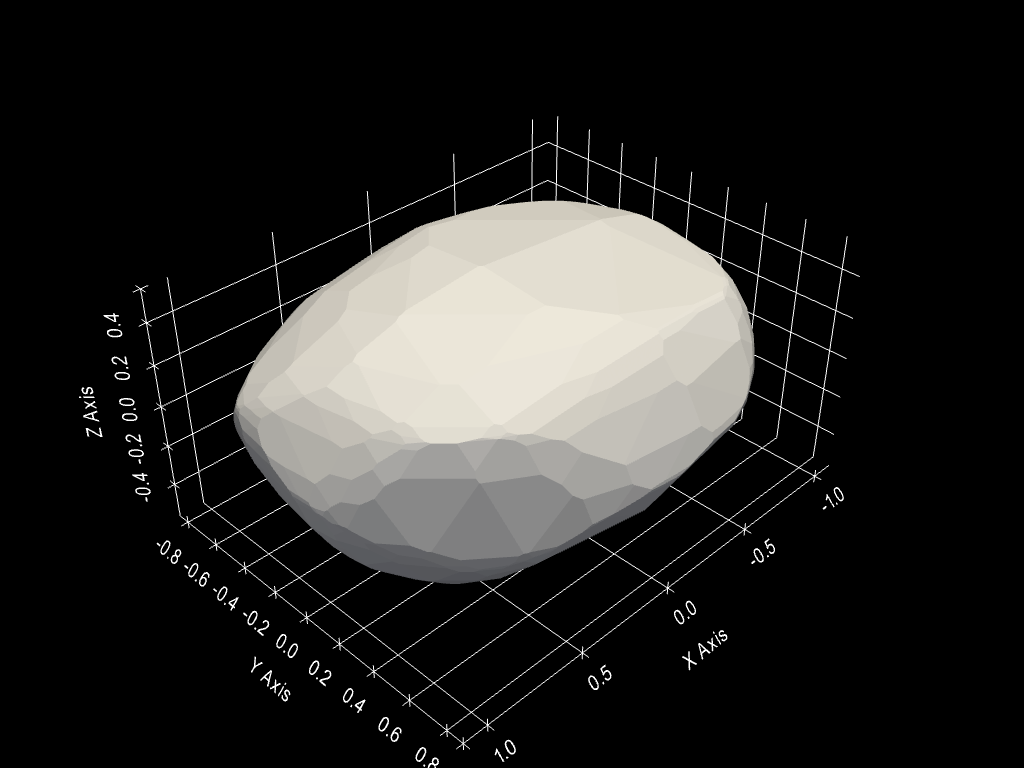} &  \includegraphics[width=0.475\linewidth]{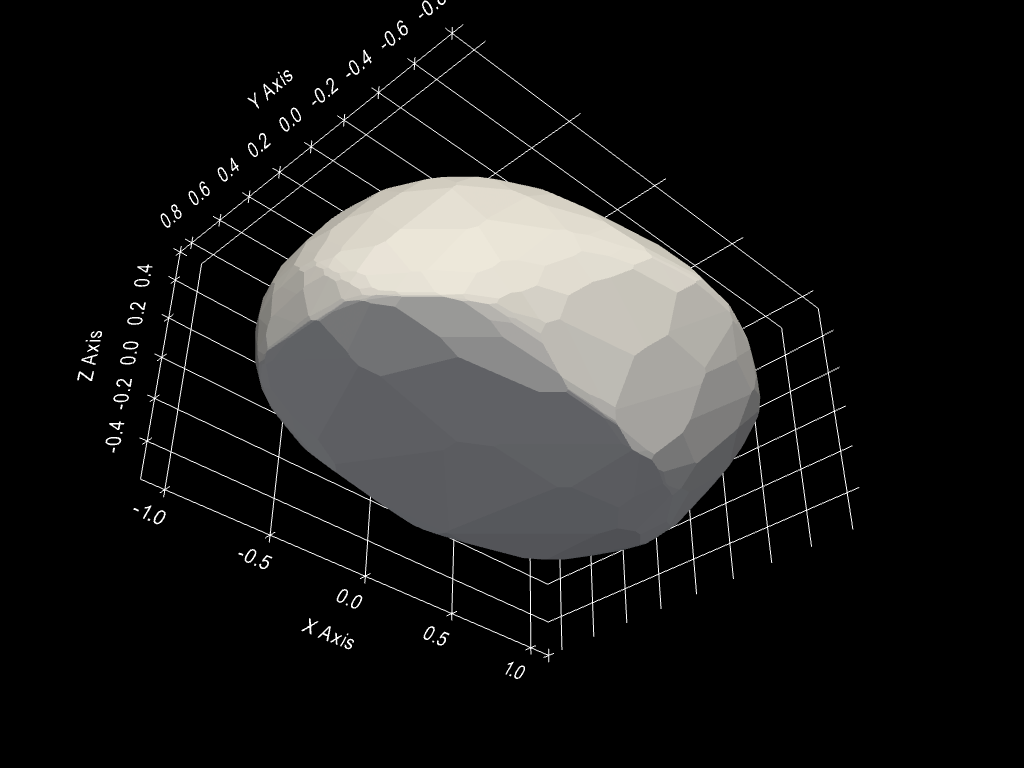}  \\
         \includegraphics[width=0.475\linewidth]{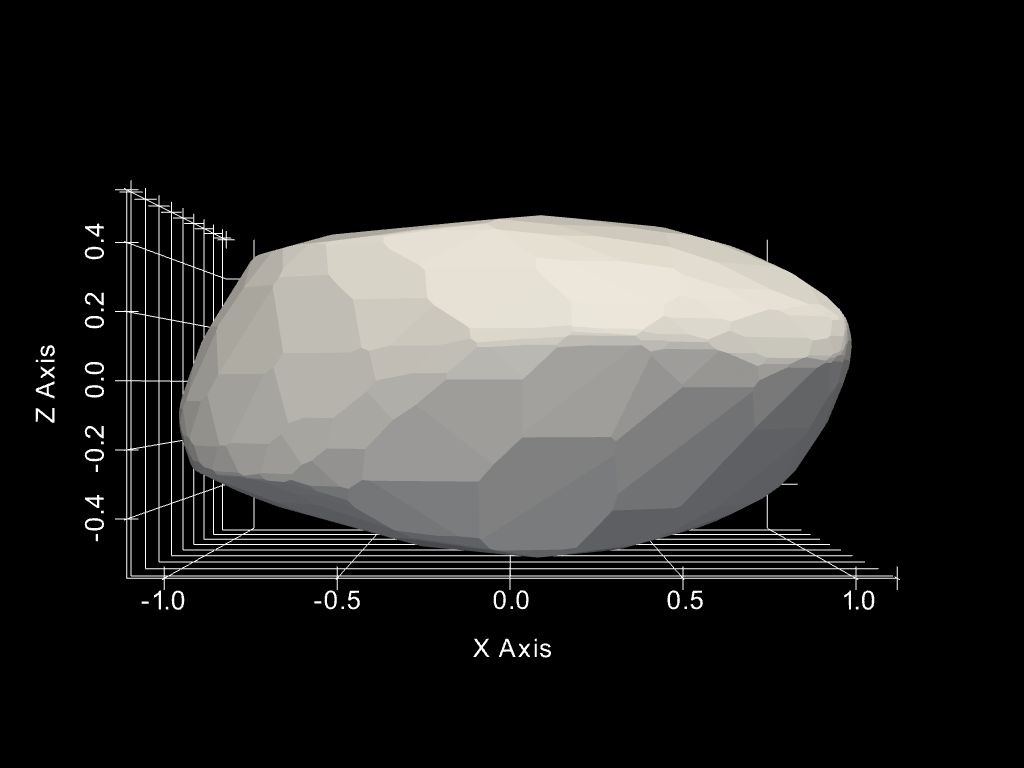} &  \includegraphics[width=0.475\linewidth]{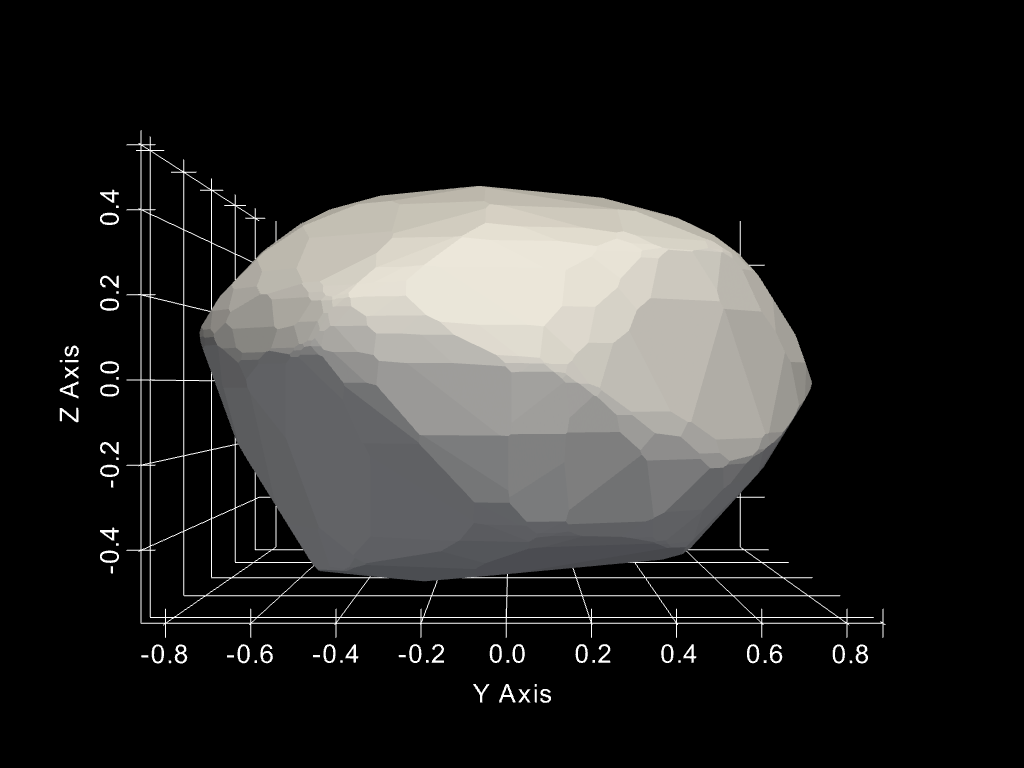}
    \end{tabular}
    \footnotesize \caption{Convex shape model of \ud{} using our preferred pole solution displayed in a variety of viewing geometries. View descriptions in left to right progression: positive isometric, negative isometric, along y-axis, and along x-axis. The scale values are unitless and show the relative sizes of the $xyz$-axes of our convex approximation. The equator of \ud{} is parallel to the $x$-$y$ plane. The light source was arbitrarily placed and does not necessarily reflect the location of the Sun for any given observation.}
    \label{fig:shapemodel}
\end{figure*}

\begin{figure*}
    \centering
    \begin{tabular}{ccc}
        \includegraphics[width=0.3\linewidth]{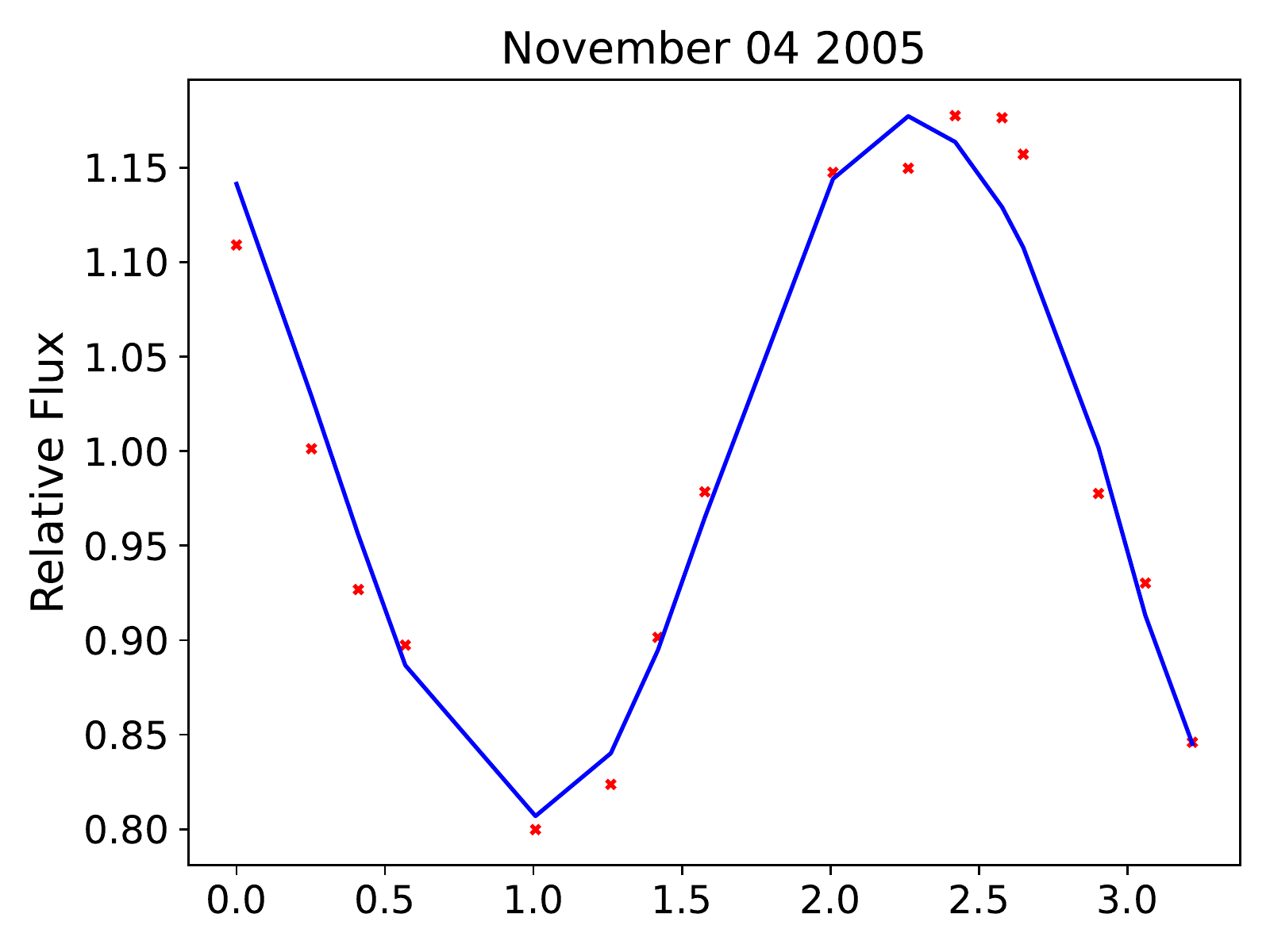} &  \includegraphics[width=0.3\linewidth]{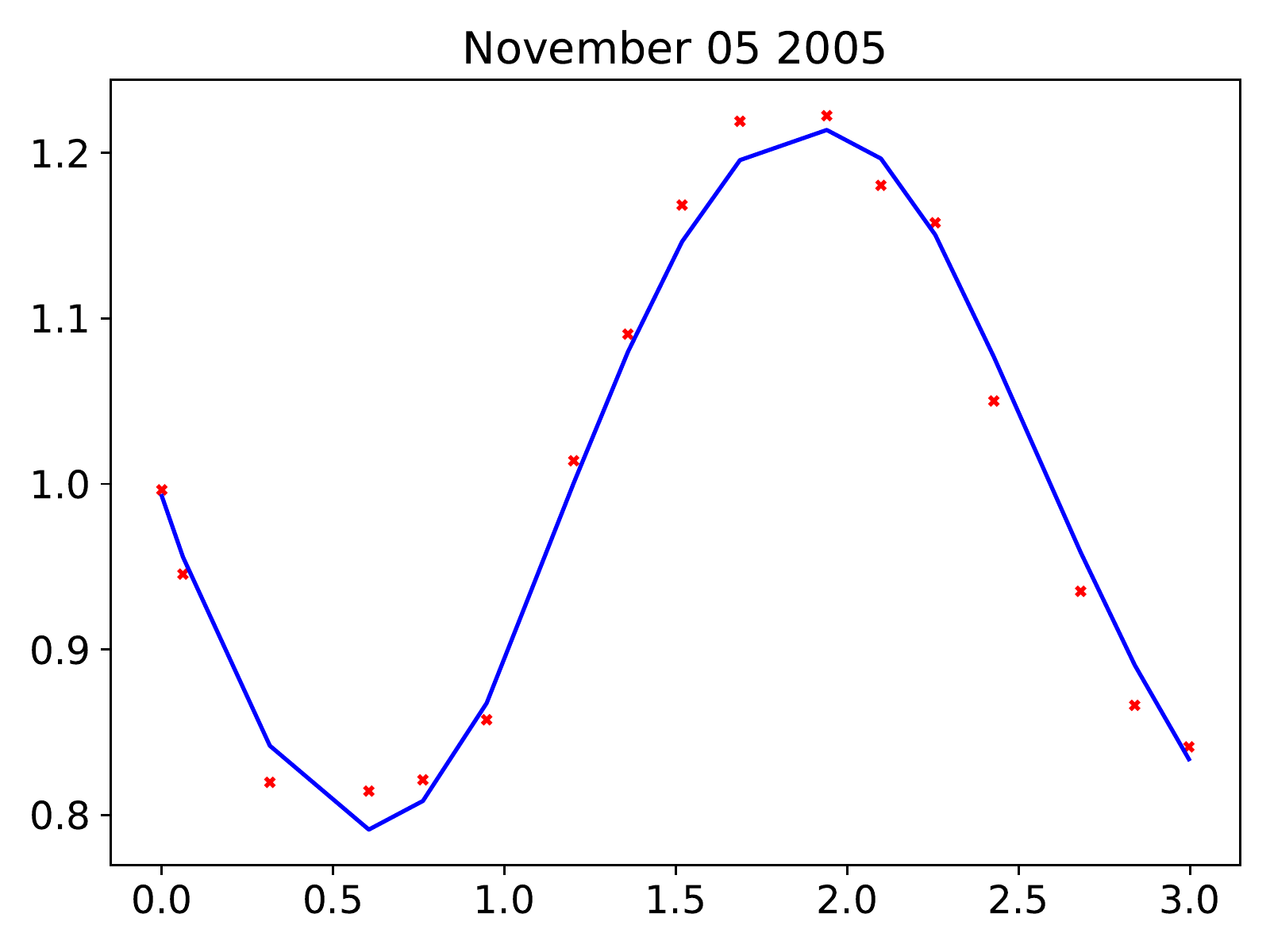} 
         & \includegraphics[width=0.3\linewidth]{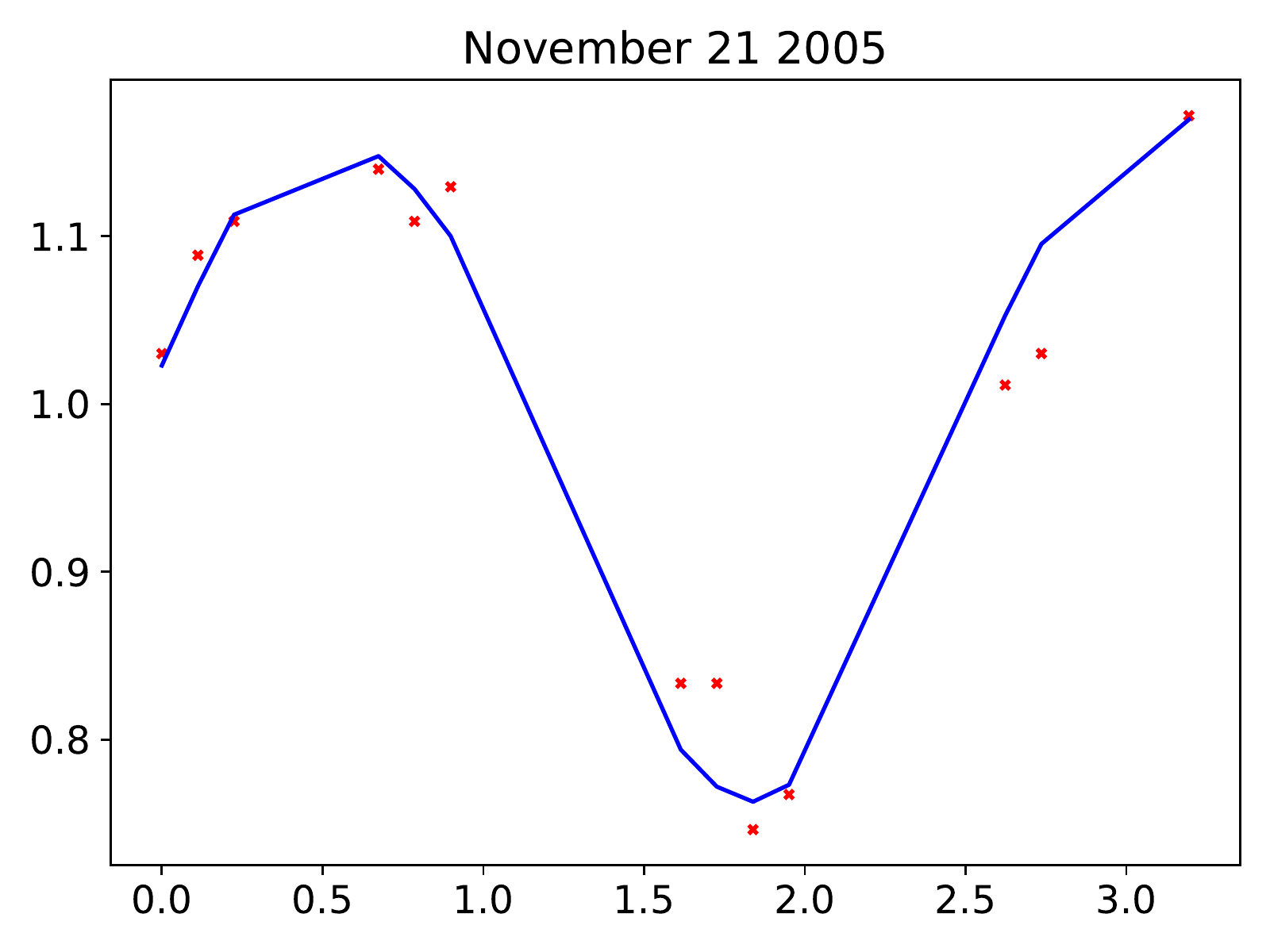} \\
         \includegraphics[width=0.3\linewidth]{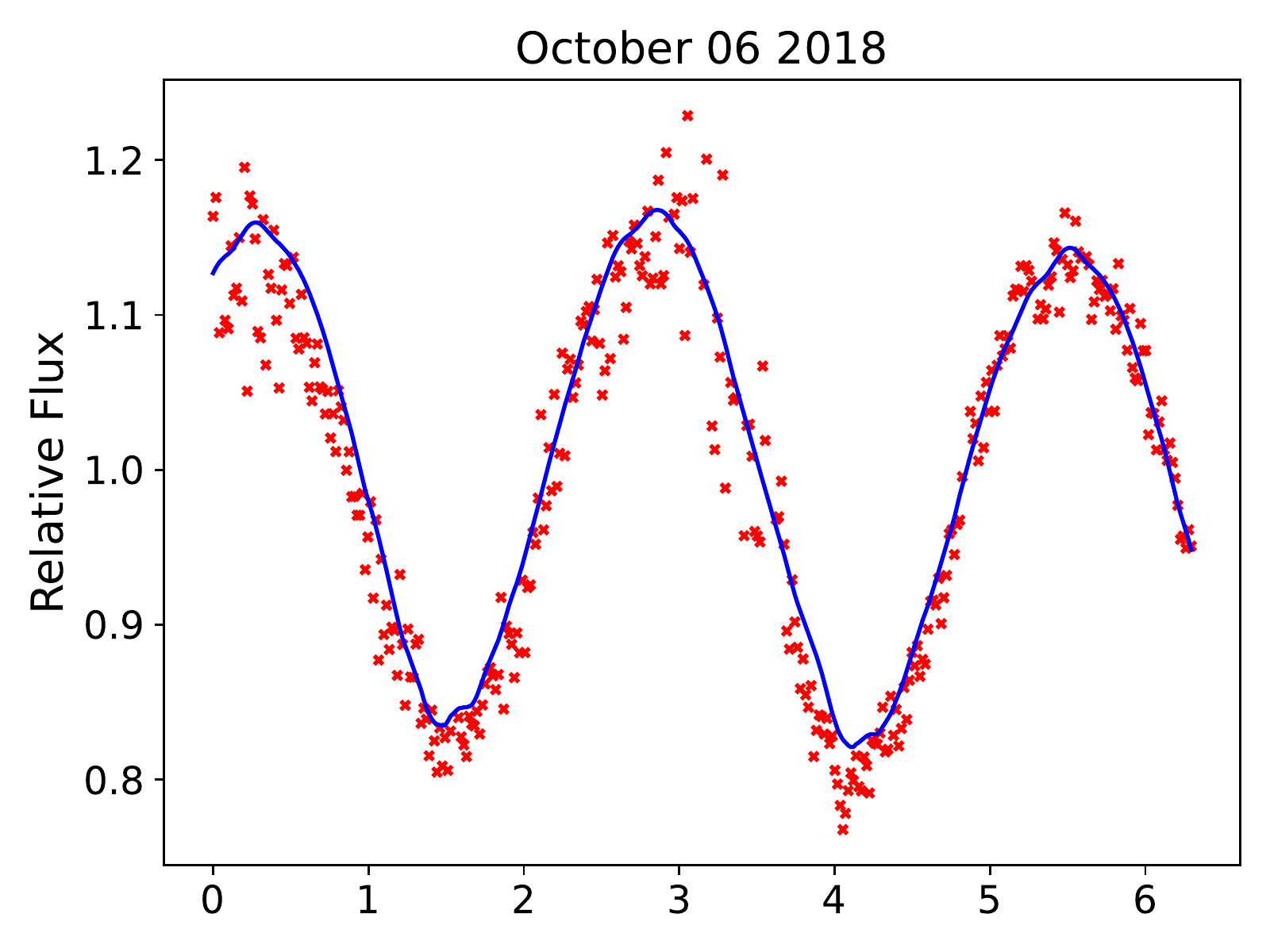} &  \includegraphics[width=0.3\linewidth]{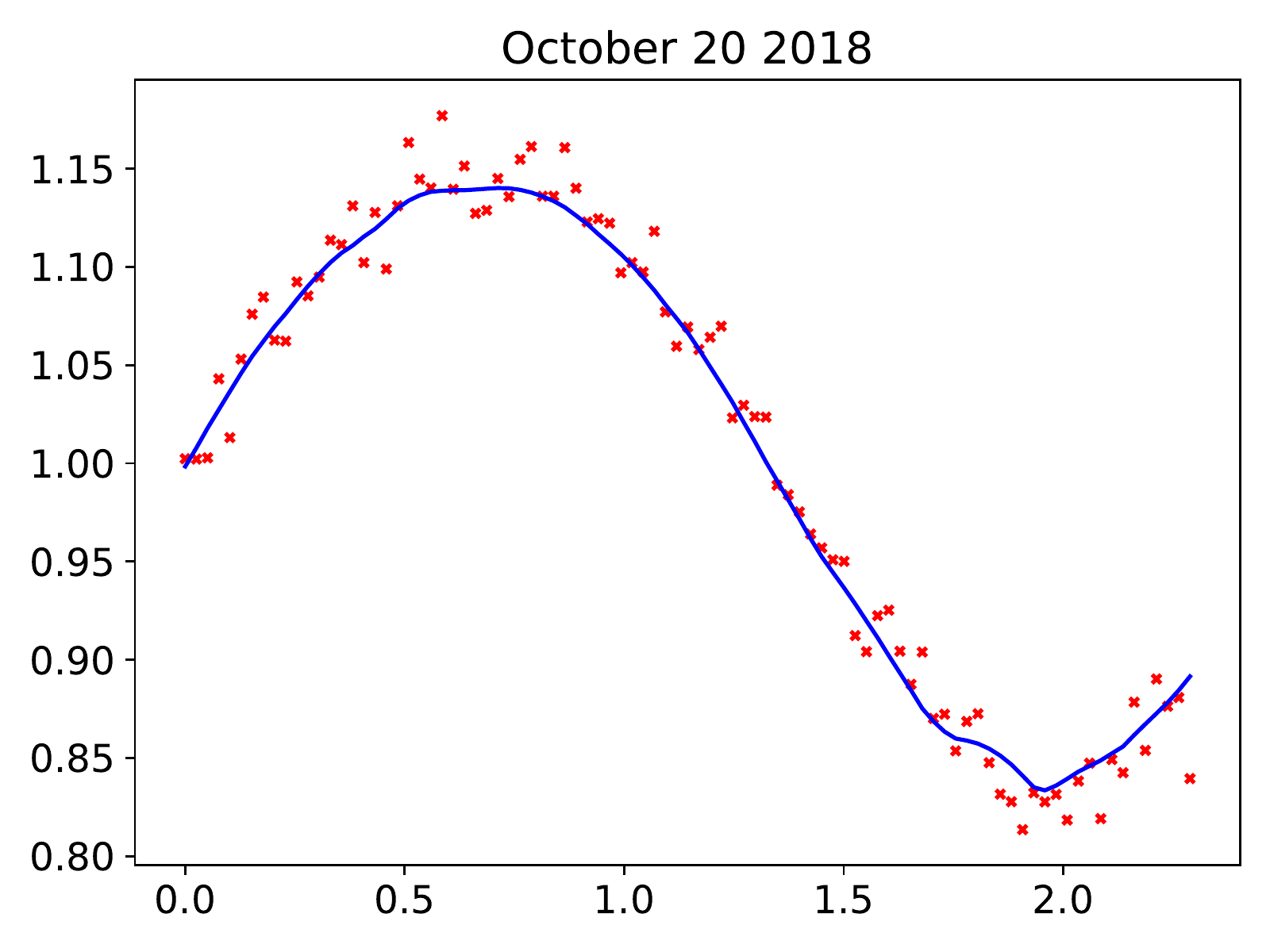} 
         & \includegraphics[width=0.3\linewidth]{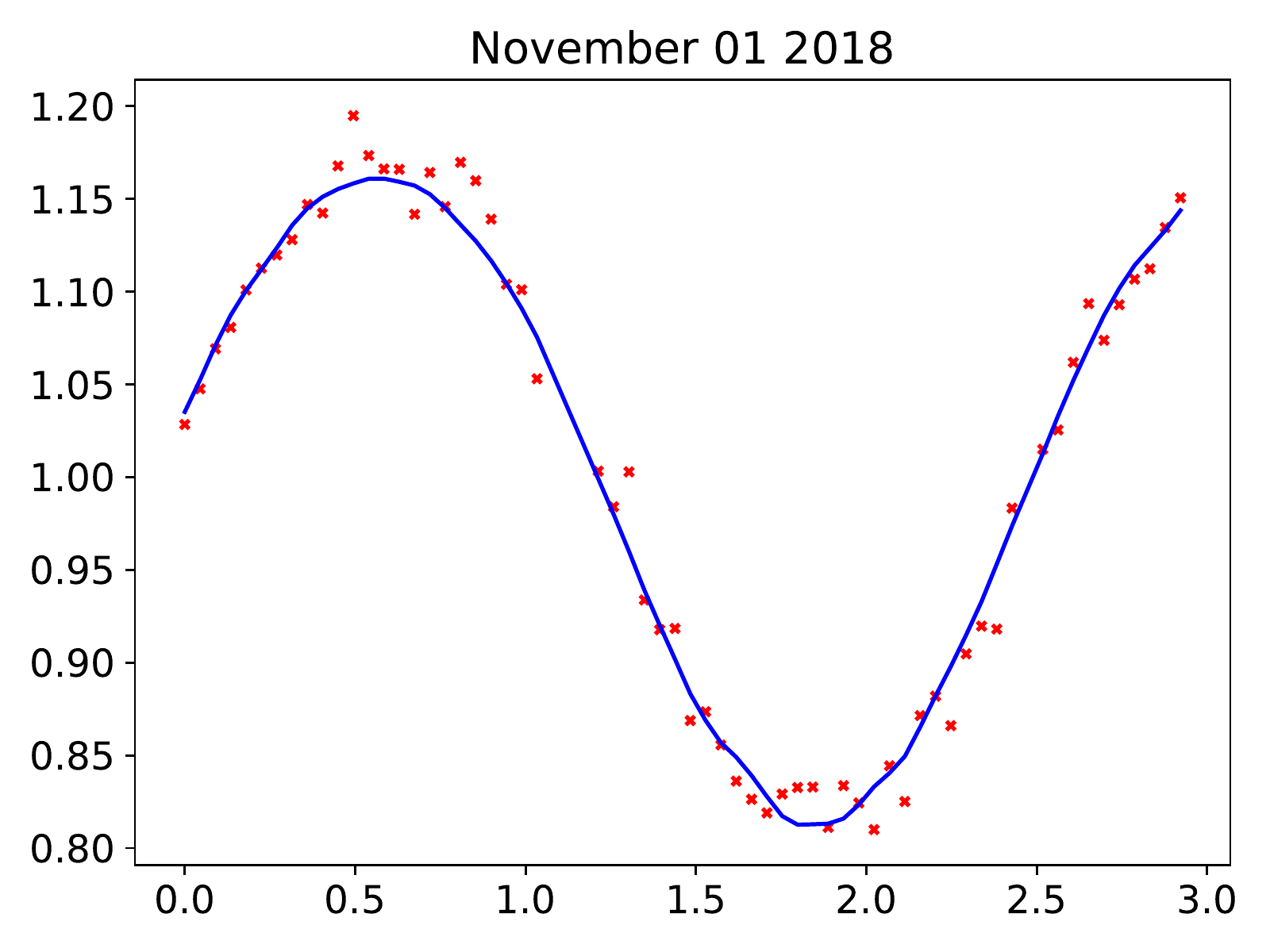} \\
         \includegraphics[width=0.3\linewidth]{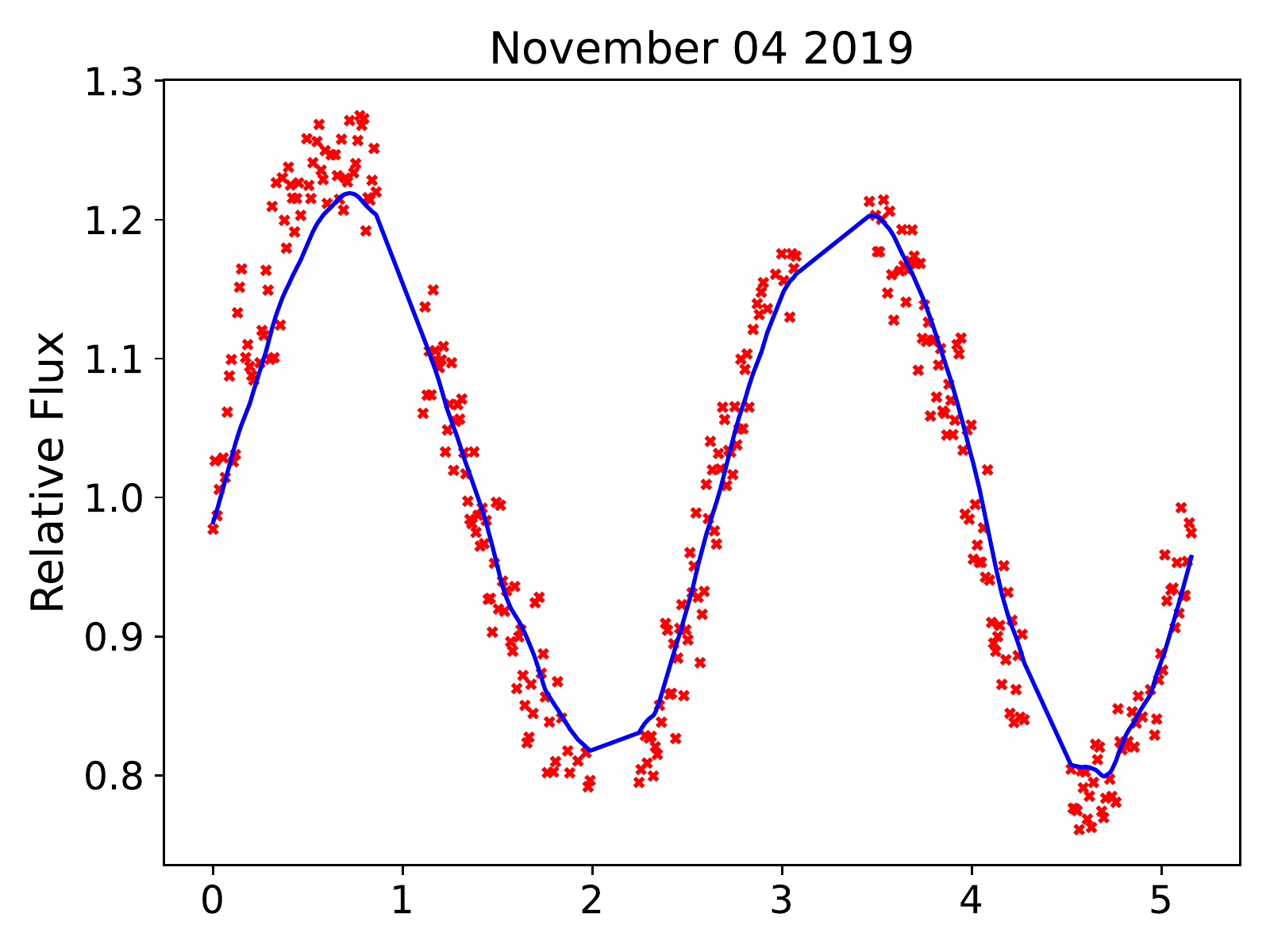} &  \includegraphics[width=0.3\linewidth]{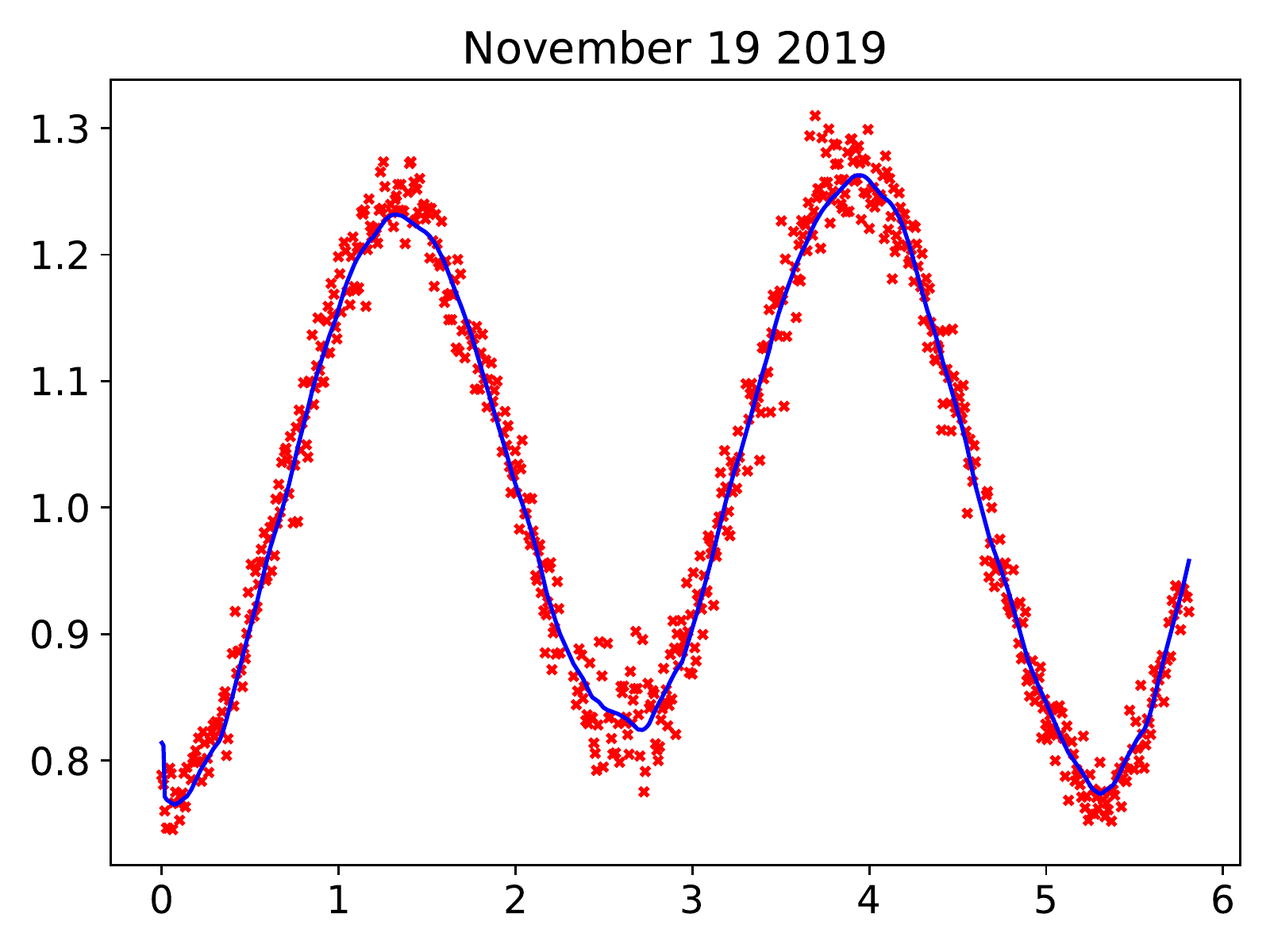} 
         & \includegraphics[width=0.3\linewidth]{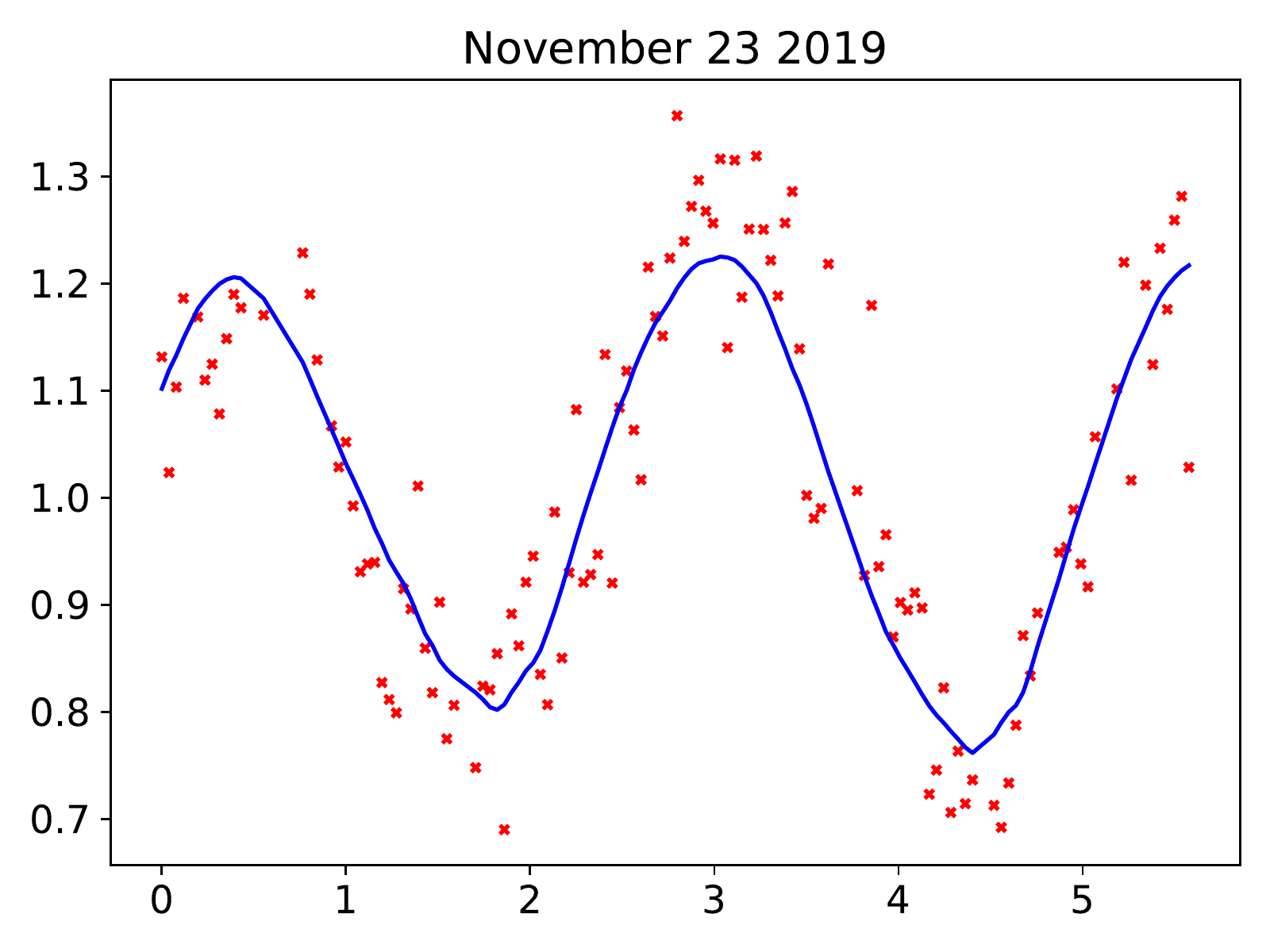} \\
         \includegraphics[width=0.3\linewidth]{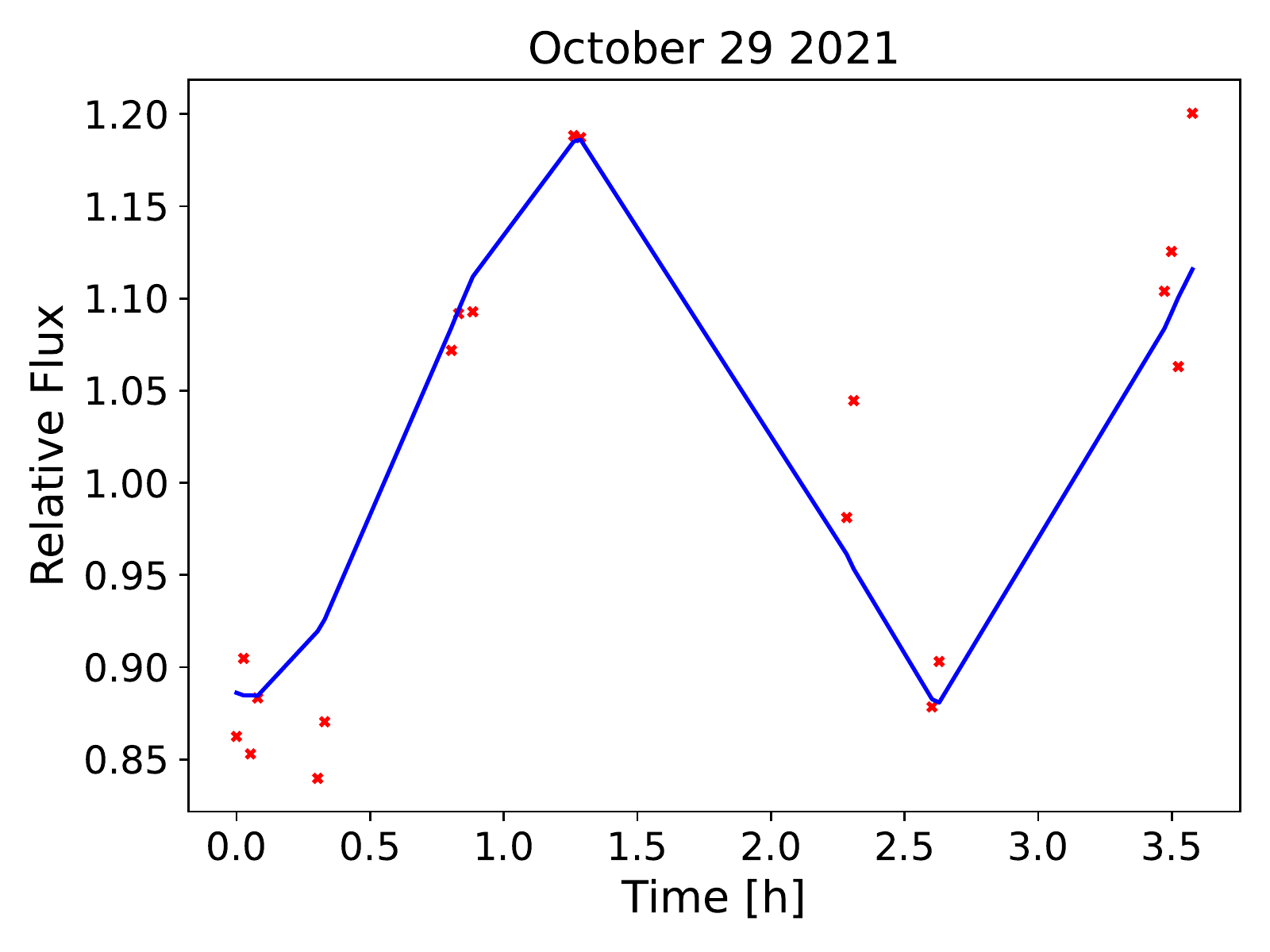} &  \includegraphics[width=0.3\linewidth]{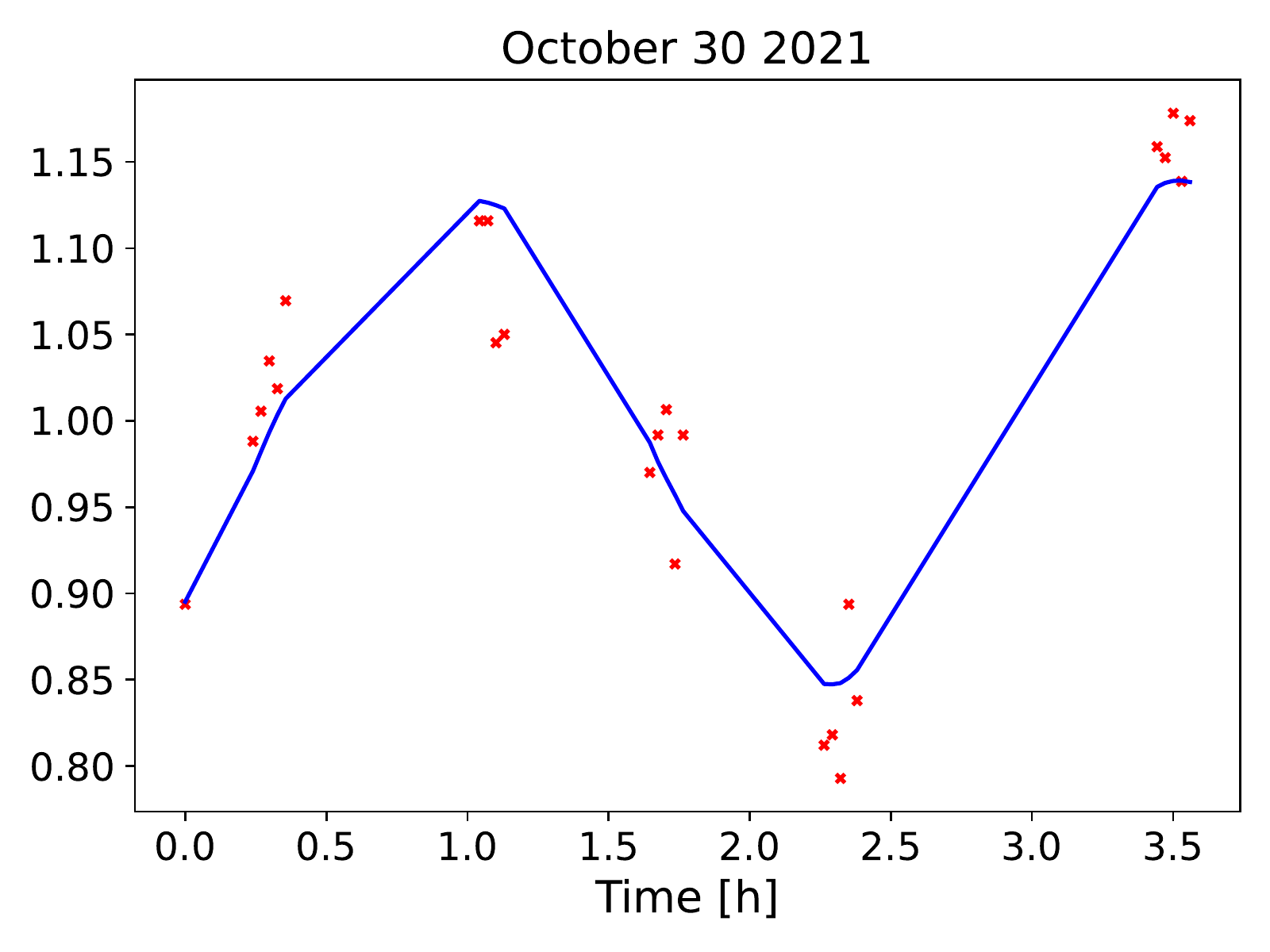} 
         & \includegraphics[width=0.3\linewidth]{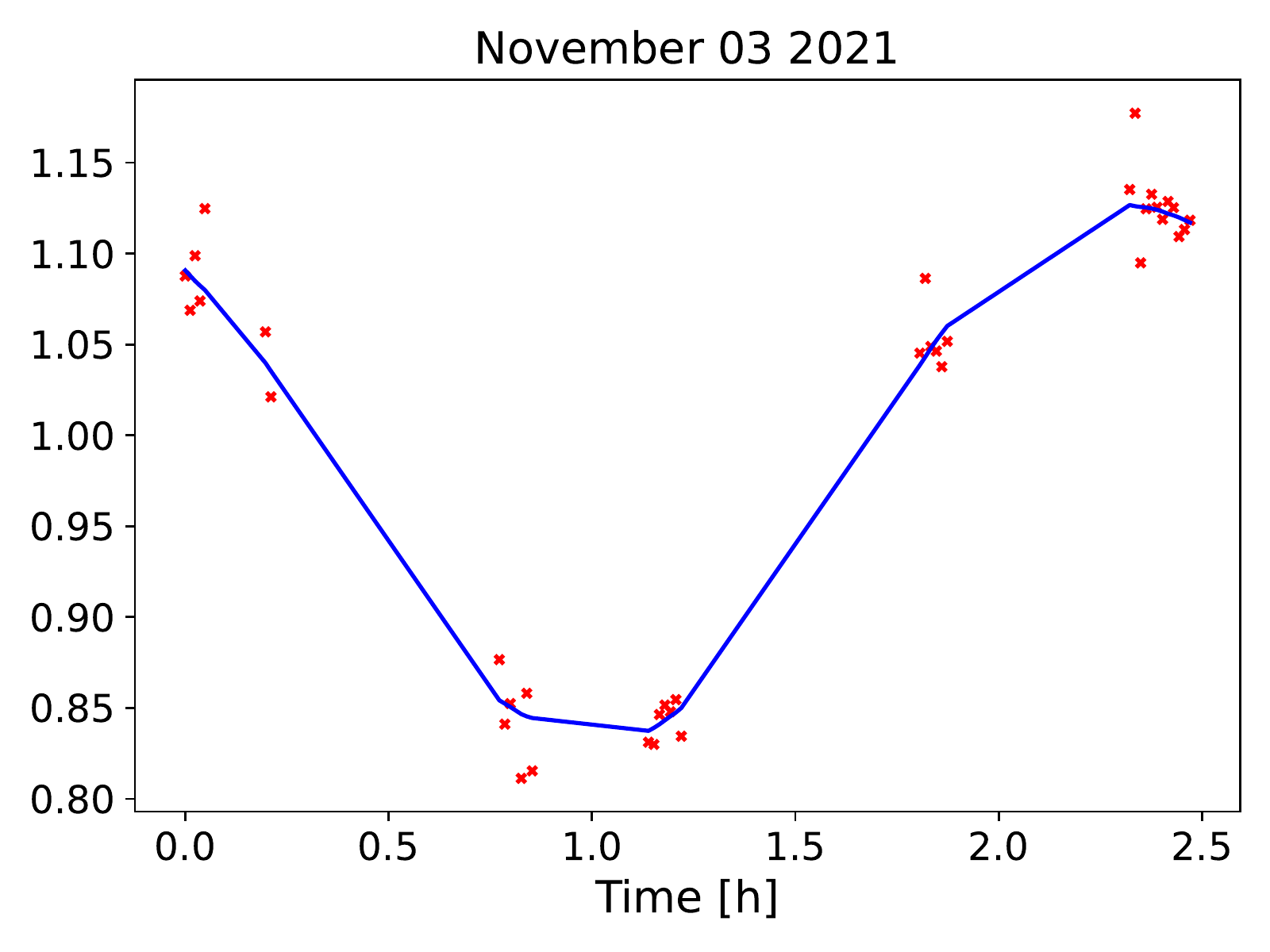} \\
    \end{tabular}
    \footnotesize \caption{Suite of \ud{} light curves compared to model fits. Each plot features a single light curve from assorted epochs in 2005, 2018, 2019, and 2021. Overlaid are light curve predictions of our preferred model from the convex inversion procedure represented as a solid blue line. UT dates marking the start of observations for each light curve are present as plot titles. The data behind this figure are available and include all 84 individual observations; see Tables \ref{Table:UDobs1}, \ref{Table:UDobs2}, and \ref{Table:UDobs3}. The data is in the form of a DAMIT input file. User can run the code to get the model values or extract the observed, relative brightness from the input file.}
    \label{fig:plots2019}
\end{figure*}

The convex shape (Figure \ref{fig:shapemodel}) from our preferred pole solution  $(\lambda_p,\beta_p) = (116\fdegree6, -53\fdegree6)$ is nearly identical to our other candidate solution $(\lambda,\beta) = (300\fdegree3, -55\fdegree4$). This is expected because there remains a $180\ddeg$ ambiguity in ecliptic longitude for the two solutions for roughly the same pole latitude. We thus expect features to be mirrored across the xz-plane. As an additional test in the validity of these two solutions, we computed light curve predictions of these two shapes and analyzed them by eye to find equally good agreement with all of our light curves. A suite of twelve plots illustrating the fits between synthetic and real light curves from epochs in 2005, 2018, 2019, and 2021 using our preferred solution is shown in Figure \ref{fig:plots2019}.

\section{Current State of Knowledge of \ud{} Physical Properties}
\label{sec:physical}

For reference, we provide a master summary of known properties for \ud{} in Table \ref{Table:UDparams1}.

\label{sec:heterogeneity}
\begin{figure*}
    \centering
	\begin{tabular}{cc}
    \includegraphics[width=0.455\linewidth]{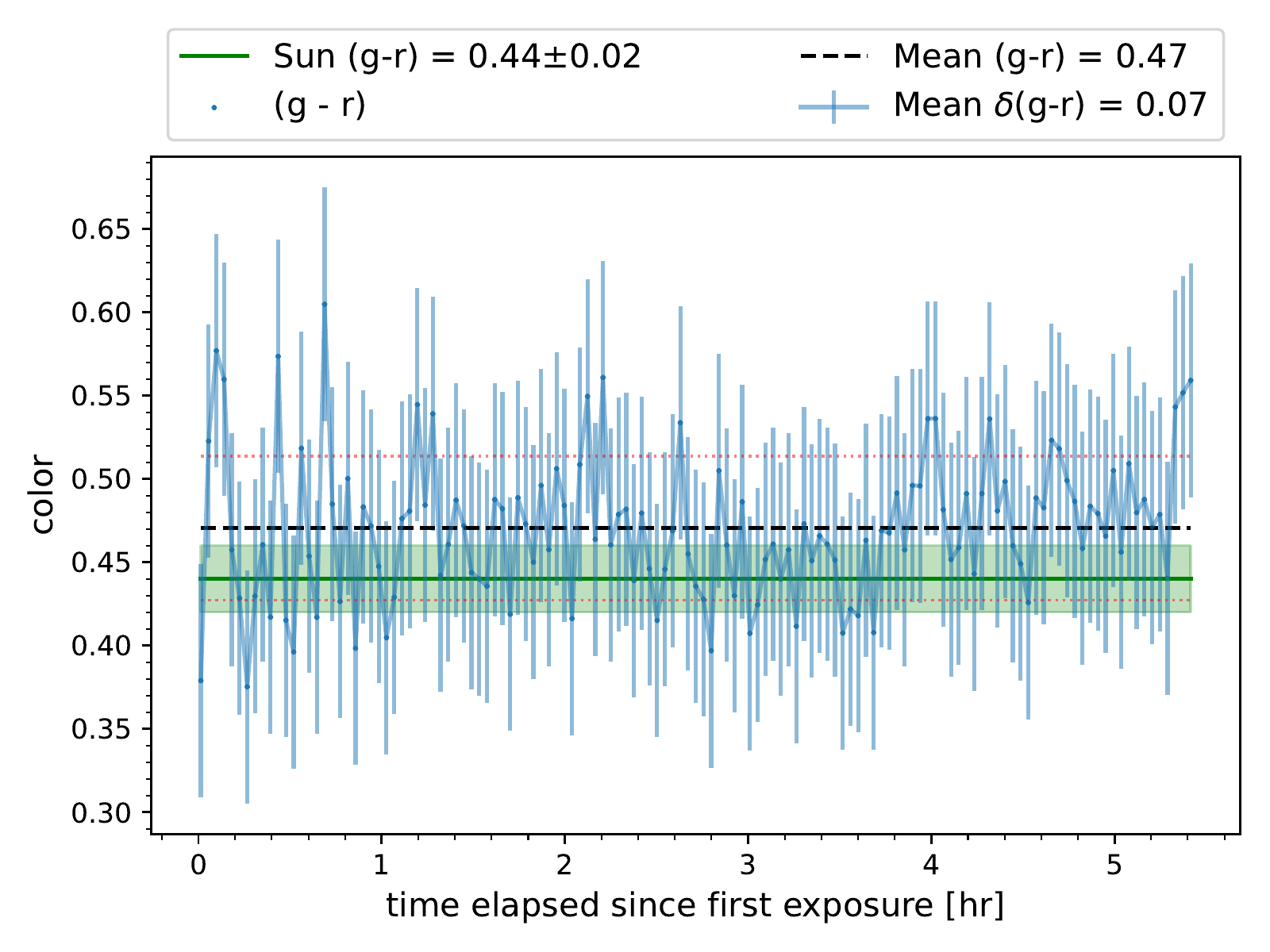} & \includegraphics[width=0.455\linewidth]{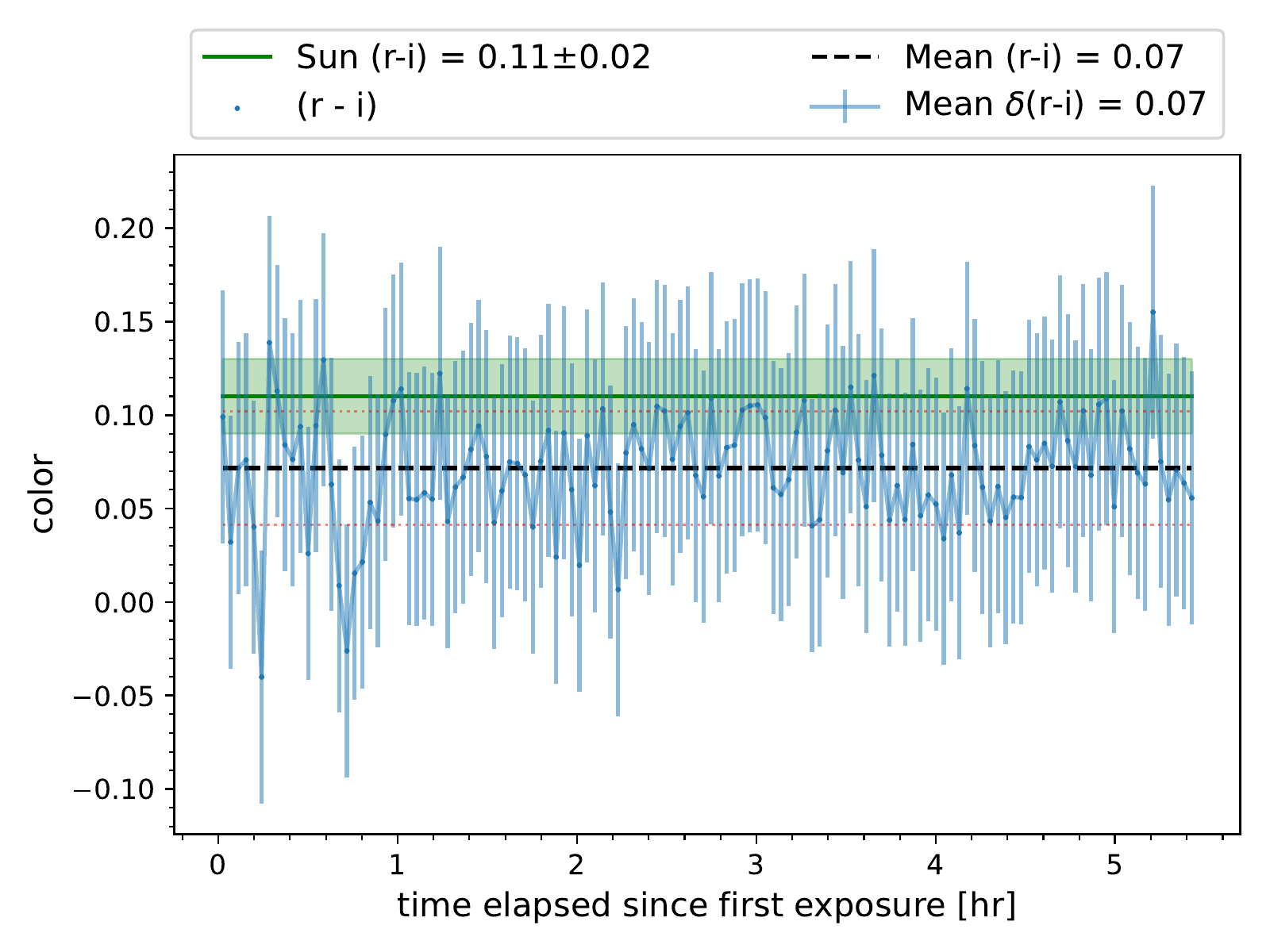}
	\end{tabular}
	\caption{\textbf{Left:} ($g - r$) color as a function of rotation. The data points were computed by subtracting linearly interpolated $r$ from measured $g$ for a given $r$-$g$-$r$ sequence. The calculated mean of these residuals is shown as the dashed horizontal black line. We represent one standard deviation from the mean as the dotted horizontal red lines, which confirm no signs of color variation outside of this limit. The ($g - r$) color of the Sun is underlaid as the green horizontal line with surrounding error for reference.  \textbf{Right:} Same as the left but with ($r - i$) color.}
	\label{fig:residuals}
\end{figure*}

\subsection{Albedo, Size, and Shape}

Thermal emission observations, coupled with photometry, are of particular value for albedo and size determination. A thermophysical study of \ud{} by \cite{masiero_thermophysical_2019} presented a value of $0.14 \pm 0.09$ for the geometric albedo and $1.2 \pm 0.4$ km for the effective diameter. However, \cite{masiero_thermophysical_2019} note that the modeling was only based on data from two observational epochs from the NEOWISE survey.

The most recent values to be reported come from \cite{devogele_new_2020}, with an effective diameter of $1.3 \pm 0.2$ km and geometric albedo of $0.10 \pm 0.02$ based on photometry and an albedo-polarimetry relation \citep{cellino_calibration_2015}. \cite{devogele_new_2020} also included thermophysical modeling  (using a different technique from that used in \citealt{masiero_thermophysical_2019}) to derive a geometric albedo of $0.14 \pm 0.07$ and effective diameter of $1.12^{+0.49}_{-0.21}$ from the same limited NEOWISE data set. Despite the slight differences in these nominal values, they are all consistent and suggest that the albedo and diameter of \ud{} are well constrained.

Our shape models of \ud{} reveal that the general shape can be inferred as a triaxial ellipsoid with $a/b = 1.41$ axis ratio that produces symmetric sinusoidal light curves with amplitude of about 0.4 mag. We notice prominent flat areas at the north and south poles on our convex shape approximation, which is expected, as these areas are the least constrained with unresolved disk-integrated photometry. \cite{devogele_new_2020} is the only other work to have attempted shape modeling of \ud{} using our methods (as presented in \citealt{kaasalainen_optimization_2001}), and they showed that their light curve data alone could not eliminate the possibility of a sidereal period of $\sim7.85$ hr. With the addition of our new data, the possibility of a three-peaked light curve and consequently a shape more akin to a tetrahedron is nonviable.

\subsection{Surface Colors}

\begin{table}
\caption{Mean Surface Colors of \ud{} and Phaethon.}
\raggedright
\small
\begin{tabular}{lcccc}
  \hline\hline    Body   & $B - V$       & $V - R$        & $R - I$       & Ref.             \\ \hline 
2005 UD  & $0.65 \pm 0.02$ & $0.32 \pm 0.01$ & $0.32 \pm 0.02$  & 1          \\
2005 UD  & $0.63 \pm 0.01$ & $0.34 \pm 0.01$  & $0.30 \pm 0.01$ & 2     \\
2005 UD  & $0.66 \pm 0.02$ & $0.35 \pm 0.02$  & $0.33 \pm 0.02$ & 3        \\
Phaethon & $0.64 \pm 0.02$ & $0.31 \pm 0.02$  & $0.31 \pm 0.03$ & 4           \\
Phaethon & $0.61 \pm 0.01$ & $0.34 \pm 0.03$  & $0.31 \pm 0.03$ & 5 \\
Phaethon & $0.59 \pm 0.01$ & $0.35 \pm 0.01$  & $0.32 \pm 0.01$ &  6       \\
Phaethon & -           & 0.34         & -           & 7  \\      \hline
\end{tabular}
\vspace{2pt}

\raggedright
\footnotesize
\textbf{References.} (1) this work; (2) \citealt{kinoshita_surface_2007};  (3) \citealt{jewitt_physical_2006}; (4) \citealt{lee_investigation_2019}; (5) \citealt{kasuga_observations_2008}; (6) \citealt{dundon_enigmatic_2005}; (7) \citealt{skiff_near-earth_1996}.
\label{tab:colors}
\end{table}

We analyzed a time series of photometric colors of \ud{} derived from $341\times30$ s exposures taken in a $g$-$r$-$i$ sequence at the NOT (Section \ref{sec:obs}) to investigate previously reported variations in surface color \citep{kinoshita_surface_2007}. These multifilter light curves span one complete rotation of the body. Calibrated photometry using the Aperture Photometry Tool (APT; \citealt{laher_aperture_2012}) shows the following apparent magnitudes: $g=19.34\pm0.01$, $r=18.87\pm0.01$, and $i=18.80\pm0.01$. We thus report colors as $(g-r) = 0.47\pm0.01$ and $(r-i) = 0.07\pm0.01$. To compare with Johnson--Cousins colors of \ud{} from prior analyses, we performed color transformations using $(B-g)$, $(V-g)$, $(R-r)$, and $(R-I)$ equations from \cite{jordi_empirical_2006}. Our results are consistent with previous works and are displayed in Table \ref{tab:colors}. 


To check for rotational color variability, we computed color as a function of rotation by linearly interpolating between adjacent points, e.g. measured $g$ minus interpolated $r$ (Figure \ref{fig:residuals}). Within the signal-to-noise ratio of our data we did not detect any systematic color variations that are more significant than $1 \sigma$ away from the mean across the region of the body visible during this observation period. This null detection is consistent with the results of (1) \cite{devogele_new_2020} who used separate spectroscopic and polarimetric methods to probe for surface heterogeneity, and (2) \cite{kareta_investigating_2021}, who saw no color variations in the near-infrared.

Our above-mentioned results are in contrast to \cite{kinoshita_surface_2007}, who reported  $\sim0.2$ mag $R-I$ color variation. One explanation for this discrepancy is that \ud{}'s surface color differs as a function of latitude, as there was a $\sim52\ddeg$ difference in the subobserver ecliptic latitudes accessed by these two color data sets. This modest difference in subobserver latitude suggests that any color heterogeneity on the surface would have to be confined to small ($<50\ddeg$ in latitude) yet highly contrasting spots in order to have an influence on the hemispherical averages represented by unresolved, ground-based photometry. Photometric modeling of such a spotted surface could provide insight on whether this interpretation is physically plausible, but this is beyond the scope of this work. Notably, a recent study by \cite{maclennan_evidence_2022} found evidence that Phethon's surface is heterogeneous as a function of latitude, which adds merit to the above hypothesis. Future multicolor photometric observations could provide further insight into \ud{}'s surface properties.

\begin{figure*}
    \centering
    \includegraphics[width=1.0\columnwidth]{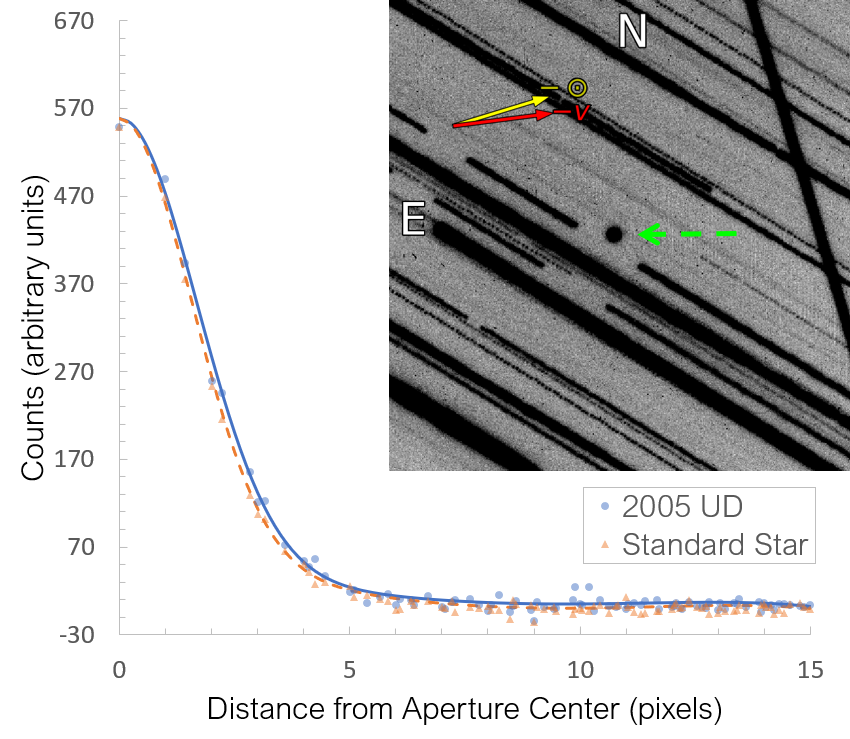}
    \includegraphics[width=1.0\columnwidth]{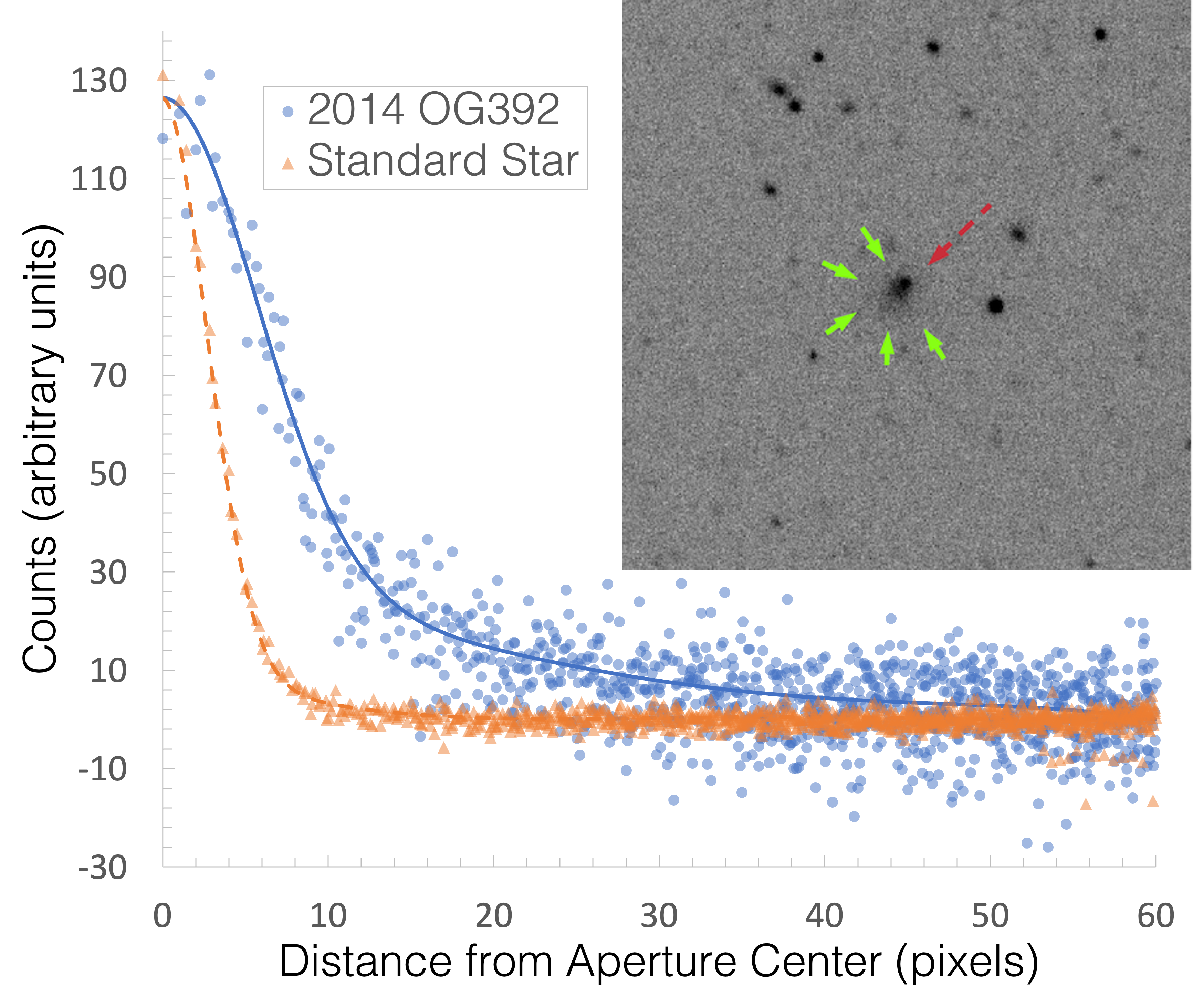}
    \caption{Left: radial surface brightness profiles of \ud{} and a nearby field star in an LDT stacked image from UT 2019 November 18. We see no indication of extended emission or faint activity. The stacked image of the asteroid corresponds to $\sim$10,960~s (about 3 hr) of integration time and reaches a limiting magnitude of about 26.8~mag arcsec$^{-2}$. The stellar profile we measured was from a single 16 s exposure, and we normalized it to the peak of stacked asteroid profile to facilitate comparison. Right: similar plot from \cite{chandler_cometary_2020} featuring active Centaur C/2014 OG392 (Pan-STARRS) to demonstrate the profile structure for an active object (reproduced with permission from the authors).}
    \label{fig:stackprofile}
\end{figure*}

\subsection{Activity and Dust Production Limits}
\label{sec:activity}

Given the detection of activity associated with Phaethon, we conducted a search for indications of a faint coma or dust tail around \ud{}. We utilized the image processing tool Siril\footnote{\url{https://www.siril.org/}} to stack LDT images from the night of 2019 November 18 using an average combination procedure with no rejection (Figure \ref{fig:stackprofile}). The stacked image corresponds to 10,910~s (about 3 hr) of integration time with a calculated depth of 26.8 mag arcsec$^{-2}$ within an annulus extending $\theta=1\farcsec8$--$3\farcsec6$ from the object center; this is the deepest existing stacked image of \ud{} at the time of writing. We used the APT \citep{laher_aperture_2012} to measure a radial surface brightness profile to compare this stacked image of \ud{} with a close field star to search for indications of a faint coma. We then fit the model 

\begin{equation*}
    S(r) = A + Br + Cr^2 + Dr^3 + Er^4 + Fe^{-\frac{r^2}{2\sigma}}
\end{equation*}

which is explained in \cite{laher_aperture_2012}. Next, we subtracted the background levels from the profiles and associated models before normalizing the field star radial profile to the peak flux of the stacked \ud{} profile. The results of these efforts are showcased in Figure \ref{fig:stackprofile} alongside a similar analysis performed on the active Centaur 2014~OG392 from \cite{chandler_cometary_2020} for comparison.

An additional quantitative search for a faint dust stream or tail was performed by analyzing summed pixel values in annular slices around the stacked asteroid (as demonstrated in \citealt{chandler_recurrent_2021}) to search for excess flux in the anti-solar and anti-motion direction. Along with visual scrutiny, no indications of activity from \ud{} were detected. Importantly, if \ud{} exhibits activity via thermal fracturing of the surface regolith (akin to Phaethon), we would expect difficult detection circumstances \citep{ye_deep_2021} because \ud{} was close to aphelion (heliocentric distance of 1.7 au) at the time these data were gathered. Targeted observations of \ud{} near perihelion might offer a better chance at detecting activity.

One way to place an upper limit on mass loss is by using a simple model with knowledge of the limiting magnitude within a projected annulus. We again used the stacked image from the LDT for this procedure using an annulus of size $\theta=1\farcsec8$--$3\farcsec6$. Using the procedure outlined in Section 3.2 of \cite{jewitt_properties_2013}, we derive an upper limit of either $0.37$~kg s$^{-1}$ or $0.04$~kg s$^{-1}$ based on assumed grain radii of $10$ mm (the upper limit of surface grain size from \citealt{devogele_new_2020}) and 1 $\mu$m respectively.  Additionally, we assume a material density of $1500$~kg m$^{-3}$. Following \cite{jewitt_properties_2013} we use particle ejection velocities of 1 m s$^{-1}$ for millimeter-sized grains and 1 km s$^{-1}$ for micron-sized grains. These upper limit results are largely consistent with the $\lesssim0.1$ kg s$^{-1}$ estimate derived by \cite{kasuga_wiseneowise_2022} using NEOWISE observations.

\section{A Common Origin with Phaethon?} 
\label{sec:discuss}

With additional data from new viewing geometries, we were able to constrain two possible spin solutions for \ud{}: $(\lambda_p=116\fdegree6, \beta_p=-53\fdegree6)$ and $(\lambda=~300\fdegree3, \beta=-55\fdegree4)$. Whether the presence of a mirror solution is a consequence of \ud{}'s unusual orbit or by coincidence is not exactly known. Observations of \ud{} during upcoming apparitions may yield further information from more viewing geometries so as to further constrain a single pole solution; Figure \ref{fig:obs} displays both past and upcoming viewing geometries that will be accessible in the next few years. The plausibility and implications for formation for each pole solution are discussed in the next section.

\subsection{Comparing Pole Solutions}

An intriguing result to come out of our analysis is a well-fitting pole solution that is consistent with Phaethon's in both longitude and latitude (see Section \ref{sec:pole}). This encourages a formation scenario where \ud{} separates from parent body Phaethon under a mechanism that does not disturb the spin vector. Formation via Yarkovsky--O'Keefe--Radzievskii--Paddack (YORP) spin-up \citep{rubincam_radiative_2000} is one such possible formation mechanism for the Phaethon-\ud{} pair. Certainly this is a common mechanism for the formation of most known asteroid pair systems \citep{pravec_formation_2010}, and interestingly, both the convex approximation of Phaethon from \cite{hanus_3200_2018} and radar shape model from  \cite{maclennan_evidence_2022} suggest the existence of an equatorial ridge and top-like shape, characteristic of YORPoids (\citealt{ostro_radar_2006}; \citealt{busch_radar_2011}; \citealt{naidu_radar_2015}). Under a YORP-induced, post-fission scenario, \ud{} and Phaethon would have engaged in a protobinary state before disruption, and eventually evolving into the unbound asteroid pair seen today \citep{jacobson_dynamics_2011}. Due to the complexity of this process and uncertainties regarding the mass distribution and morphology of both Phaethon and \ud{}, it remains unclear whether aligned poles are expected and/or consistent with a YORP spin-up scenario. Conversely, \citealt{pravec_asteroid_2019} found that current pole orientations of asteroids in a paired system generally do not reflect the original orientations at the time of separation, possibly due to solar torques \citep{breiter_efficient_2005}, YORP effects, and/or planetary encounters. For Phaethon (and possibly \ud{}), the added dynamical effects of de-volatilization add more uncertainty on how pole orientations would be affected (discussed more below). In the case of our preferred solution, which only shares a common ecliptic latitude with Phaethon, we do not consider pole realignment due to planetary flybys a likely scenario because the minimum orbital intersection distance (MOID) value for both bodies is too high for all planets. YORP ultimately may also be invalid here due to the proposed recent separation of less than 100 kyr (\citealt{hanus_near-earth_2016}; \citealt{maclennan_dynamical_2021}), which is much shorter than the tens of millions of years required for significant changes due to YORP \citep{rubincam_radiative_2000}. To probe the extent to which current-day spin poles for \ud{} and Phaethon line up or differ as a result of any of these physical process would require detailed modeling beyond the scope of this paper.

Due to Phaethon's rare status as an active asteroid, volatile-driven separation is another possible formation mechanism for this pair. This conjecture, however, does not rule out rotational fission entirely since induced torques due to outgassing cause spin changes in comets \citep{steckloff_formation_2016}. Akin to similar processes responsible for cometary splitting (see \citealt{jewitt_d_systematics_2021} and references therein), episodes of activity in an ancestor body could have worked, perhaps in concert with YORP, to accelerate the body to its critical spin rate, resulting in fission. However, the fact that \ud{} is clearly in a principal axis rotation state (i.e. it is not tumbling) may speak to a more ordered scenario of formation that did not involve processes, like volatile-driven outgassing, that can produce nonprincipal axis rotation. Damping time estimates using Eq. 11 from \citet{pravec_tumbling_2014} of a nonprincipal axis rotator of \ud{}'s size and rotation exceed the proposed age of this system by millions of years, further suggesting that a chaotic, purely volatile-driven splitting event was unlikely.

Based on our results and arguments presented above, we lean toward a common origin due to a YORP fission scenario being the most likely progenitor for the \ud{} Phaethon cluster. This interpretation is consistent with that made by \cite{huang_photometric_2021}, who also presented a Phaethon-like pole solution for \ud{} using a different method. We encourage modeling of this specific sequence of events to probe the effects on spin pole in such a scenario.

\subsection{Nongravitational Influences}

The Yarkovsky effect (\cite{bottke_yarkovsky_2006}) is relevant when considering nongravitational perturbations for small (generally less than $\sim$10 km diameter) bodies. In general, a secular decrease in semimajor axis is expected for retrograde rotators (as opposed to outward drift for prograde rotators) such as Phaethon and \ud{}. \ud{} is approximately one-fifth the size of Phaethon, which would subject it to more rapid inward drift in semi-major axis. However, the semi-major axis of \ud{} is currently greater by $\sim3.5\times10^{-3}$ au than that of Phaethon and dynamical integrations \citep{ohtsuka_apollo_2006} suggest that that has been the case for thousands of years. Furthermore, the presumed recent ($<100$ kyr ago) separation of this pair means that the Yarkovsky effect has not had enough time to significantly alter these object's orbits. Typical Yarkovsky drift rates of $10^{-4}$ au Myr$^{-1}$ for kilometer-scale bodies suggest that $\sim10$ Myr would be needed to explain their current difference in semi-major axis, and that does not account for any time associated with a necessary change in pole orientation (e.g., by the YORP effect) that would allow for Yarkovsky drift to be in the right direction. It is additionally unlikely that the current low levels of activity seen for Phaethon play any significant role in the orbital dynamics of the system, though prior epochs with higher levels of activity may have contributed to the present-day separation.

If Phaethon and \ud{} did in fact separate recently, it is suggested \citep{maclennan_dynamical_2021} that orbital dynamics governed by interactions with the terrestrial planets remain the most plausible dynamical pathway to produce the configuration of orbital elements we see today. However, a definitive time line associated with a separation event between Phaethon and \ud{} has yet to come to fruition. We hope that improved orbit solutions and physical evidence of a fission-type disruption event following the DESTINY+ flyby could provide more insight into this issue.

Lastly, it is worth mentioning that (225416) 1999 YC is a putative third component in the \ud{}-Phaethon system (\citealt{kasuga_observations_2008}; \citealt{ohtsuka_apollo_2008}). The addition of this body to the system supports a rotational fission formation scenario \citep{hanus_3200_2018}, but a spectral type inconsistent with the other two bodies \citep{kasuga_observations_2008} and a significantly larger semi-major axis are hard to reconcile with any common origin theory.

Given these considerations, it seems unlikely that nongravitational forces would have played an important role in any evolution of \ud{} relative to Phaethon. This does not necessarily influence arguments in favor or against these two objects sharing a common origin.

\section{Summary and Future Work} 
\label{sec:summ}

We conducted observations of \ud{} during its recent apparitions in 2018, 2019, and 2021 using six different telescopes (4.3 m LDT, 2.6 m NOT, Danish 1.54 m, Ond\v{r}ejov 0.65 m, and 0.6 m TRAPPIST-North and TRAPPIST-South) motivated by a fortuitous opportunity to collect data at new aspect angles. Our goals for this analysis included finding a unique sidereal rotational period, pole solution, and thus an accurate shape for \ud{} by means of light curve inversion. We supplemented these data with archival dense and sparse light curves from epochs in 2005--2021 prior to performing light curve inversion. We presented a refined sidereal rotational period of $P_{\text{sid}} = 5.234246 \pm 0.000097$ hr. We conclude that \ud{} has two equally well-fitting spin solutions given as ($\lambda_p = 116\fdegree6$, $\beta_p = -53\fdegree6$) and ($\lambda = 300\fdegree3$, $\beta = -55\fdegree4$), the former having preference considering the proximity of all other candidate solutions from our extensive light curve inversion trials and thermophysical modeling results from \citealt{devogele_new_2020}. Furthermore, the preferred solution from \cite{huang_photometric_2021} of $(285\fdegree8\substack{+1.1 \\ -5.3}, -25\fdegree8\substack{+5.3 \\ -12.5})$ is only consistent in latitude with the latter of our two solutions given $3\sigma$ errors. Additional light curve data of \ud{} from more viewing geometries may be needed to further constrain a unique spin solution and/or resolve inconsistencies with earlier studies. We thus encourage follow-up observations in upcoming apparitions to either help solve this problem or confirm our results (see Figure \ref{fig:obs}).

An activity search using the deepest stacked image of \ud{} at the time of writing revealed no presence of dust production at orbital longitudes near aphelion. A simple model was used to infer an upper limit to mass loss of around 0.04--0.37~kg s$^{-1}$ depending on assumed grain size. This analysis should be applied to images of \ud{} taken at or very close to perihelion to provide further constraints on the possibility of mass loss when surface temperatures are highest.

We analyzed time-series color data of \ud{} over a full rotation and found no significant indications of surface color heterogeneity in Sloan $g$-$r$-$i$ bands. Understanding of the link between \ud{} and Phaethon could be enhanced with additional searches for color heterogeneity, as the results presented in Section \ref{sec:heterogeneity} are in contrast to those presented in \cite{kinoshita_surface_2007}, but are consistent with two other studies (\citealt{devogele_new_2020} and \citealt{kareta_investigating_2021}). Additionally, there would be value to investigating the size of a patch of surface material that would be needed to result in detectable color variation on a body the size and shape of \ud{}. This could be done, for example, through radiative transfer models or laboratory experiments.

Our secondary spin solution is aligned with Phaethon's within error, which strengthens certain common origin scenarios. One such possibility is recent separation of \ud{} and Phaethon in a YORP-induced fission event with conservation of angular momentum keeping the poles more or less aligned. However, the extent to which a parent body's spin pole orientation would be preserved between separated pieces in this circumstance is currently a question that's unable to be definitively answered. Phaethon's perihelion-driven activity complicates this thought experiment, but present-day activity episodes reveal that they may not be intense enough (\citealt{jewitt_activity_2010} and \citealt{li_recurrent_2013}) to have any real effect on spin axis.  As such, we encourage efforts to model such scenarios. 

Our spin solutions for \ud{} spark interesting implications for orbital evolution via Yarkovsky forces. Current estimates for the age of the Phaethon cluster (assuming they are genetically related) suggest that the Yarkovsky effect alone could not have resulted in the currently observed separation of members. While \citet{ohtsuka_apollo_2006} suggest dynamics consistent with a common origin, most recently \cite{ryabova_asteroid_2019} performed backward orbital integrations spanning 5000 yr and concluded that \ud{} and Phaethon do not share a common origin. It is, however, worth noting that dynamical analyses for \ud{} have thus far neglected consideration of Yarkovsky and cometary forces which could have a significant effect on the system dynamics. Although the presented spin solutions of \ud{} may help to offer clues on its ancestry, more data and a better understanding of its detailed orbital dynamics are required to fully understand the origin of this unusual object.
\vspace{-1cm}

\acknowledgments

 We would like to thank the anonymous reviewers, whose feedback improved the quality of this manuscript markedly.
 
 Support from Michael Mommert on the \textit{PhotometryPipeline} simplified data processing steps during critical phases of this project. Matthew Knight offered helpful guidance on the mass-loss modeling work in this project. We would like to give thanks to Northern Arizona University's HABLab research group, Prof. Tyler Robinson, and Prof. Chadwick Trujillo for comments and suggestions that improved this work greatly. 


The data presented here were obtained (in part) with ALFOSC, which is provided by the Instituto de Astrofisica de Andalucia (IAA) under a joint agreement with the University of Copenhagen and NOT.

This work is based on observations made with the Nordic Optical Telescope, owned in collaboration by the University of Turku and Aarhus University and operated jointly by Aarhus University, the University of Turku, and the University of Oslo, representing Denmark, Finland, and Norway, respectively; the University of Iceland; and Stockholm University, at the Observatorio del Roque de los Muchachos, La Palma, Spain, of the Instituto de Astrofisica de Canarias.

The work at Ond\v{r}ejov Observatory, the work at the Danish 1.54 m telescope on the ESO La Silla station, and the work of J.H. were supported by the Grant Agency of the Czech Republic, grant 20-04431S.

TRAPPIST is funded by the Belgian Fund for Scientific Research (Fond National de la Recherche Scientifique, FNRS) under the grant PDR T.0120.21. TRAPPIST-North is a project funded by the University of Liège, in collaboration with the Cadi Ayyad University of Marrakech (Morocco). E.J. is F.R.S.-FNRS Senior Research Associate.

\ This material is based on work supported by the National Science Foundation Graduate Research Fellowship Program under grant Nos. 2020303693 and 2018258765. Any opinions, findings, and conclusions or recommendations expressed in this material are those of the author(s) and do not necessarily reflect the views of the National Science Foundation.

J.K. and N.M. acknowledge support from NASA Hayabusa2 participating scientist grant NNX16AK68G. J.K., N.M., and M.D. acknowledge support from NASA NEOO grant NNX17AH06G awarded in support of the Mission Accessible Near-Earth Object Survey (MANOS).

M.G., L.S., and G.F. acknowledge support from the Academy of Finland.

C.K. acknowledges support by the Swiss National Science Foundation under grant 185692.

S.M. acknowledges support from the Magnus Ehrnrooth Foundation and the Vilho, Yrj\"{o}, and Kalle V\"{a}is\"{a}l\"{a} Foundation.

\ This research has made use of data and/or services provided by the International Astronomical Union's Minor Planet Center.
\ This research has made use of NASA's Astrophysics Data System.

\ This research has made use of SAO Image DS9, developed by Smithsonian Astrophysical Observatory \citep{joye_new_2003}. This work made use of the Lowell Observatory Asteroid Orbit Database \textit{astorbDB} \citep{moskovitz_astorb_2022}.
\ This work made use of the \textit{astropy} \citep{astropy_collaboration_astropy_2013} and SciPy \citep{scipy_10_contributors_scipy_2020} software packages.

\clearpage
\bibliography{references.bib}

\clearpage

\appendix


\section{Master summary table for \ud{}} 
\label{appendixA}

\renewcommand\thefigure{\thesection\arabic{figure}}
\setcounter{figure}{0}
\renewcommand\thetable{\thesection.\arabic{table}}
\setcounter{table}{0}

Table \ref{Table:UDparams1} encompasses known physical parameters for \ud{} which we include for reference.
\begin{table}
\begin{center}
\caption{\ud{} Physical Parameters}
\label{Table:UDparams1}
\begin{tabular*}{0.9\linewidth}{@{\extracolsep{\fill}}lcc}
\hline
Parameter                               & Value                                                                                             & Reference          \\ \hline
Discovery date                          & 2005-Oct-22                                                                                       & MPEC 2005-U22      \\
Discovery observer                      & E. J. Christensen                                                                                 & MPEC 2005-U22      \\
Discovery survey                        & Catalina Sky Survey                                                                               & MPEC 2005-U22      \\
\hline
& Orbital Elements & \\ \hline 
Orbit type                              & NEA                                                                                               & NASA JPL Horizons  \\
Family                                  & Apollo                                                                                            & NASA JPL Horizons  \\
Perihelion                              & 0.1627 au                                                                                         & NASA JPL Horizons  \\
Aphelion                                & 2.3867 au                                                                                         & NASA JPL Horizons  \\
Semimajor axis                          & 1.2747 au                                                                                         & NASA JPL Horizons  \\
Orbital inclination                     & $28\fdegree6677$                                                                                   & NASA JPL Horizons  \\
Mean anomaly                            & $90\fdegree7904$                                                                                   & NASA JPL Horizons  \\
Argument peri.                          & $207\fdegree597$                                                                                   & NASA JPL Horizons  \\
Eccentricity                            & 0.8723                                                                                            & NASA JPL Horizons  \\
Long. ascending                         & $19\fdegree7247$                                                                                  & NASA JPL Horizons  \\
T\_J                                    & 4.507                                                                                             & NASA JPL Horizons  \\
Orbital period                          & 1.44 yr                                                                                          & NASA JPL Horizons  \\
\hline
& Spectroscopy & \\ \hline 
Taxonomy (Bus-DeMeo)                    & B-type                                                                                            & \cite{devogele_new_2020}      \\

Spectral Slope (Optical)                         & $20\% \pm 10\%$ $\mu$m$^{-1}$                                                                       & \cite{devogele_new_2020}      \\
Spectral Slope (NIR)                         & $6\% \pm 1\%$ $\mu$m$^{-1}$                                                                       & \cite{kareta_investigating_2021}      \\
\hline
 & Thermophysics & \\ \hline 
Thermal inertia                         & $300\substack{+120 \\ -110}$ J m$^{-2}$ K$^{-1}$ s$^{-1/2}$ & \cite{devogele_new_2020}      \\
Grain size                              & 0.9-10 mm                                                                                           & \cite{devogele_new_2020}      \\
\end{tabular*}
\end{center}
\end{table}

\clearpage
\setcounter{table}{0}

\begin{table}
\begin{center}
\caption{\ud{} Physical Parameters}
(Continued)
\begin{tabular*}{0.9\linewidth}{@{\extracolsep{\fill}}lcc}
\hline
Parameter                               & Value                                                                                             & Reference          \\ \hline
Mass loss rate                          & 0.04--0.37~kg s$^{-1}$                                                                                          & This work (Section \ref{sec:activity})       \\
\hline
 & Photometry & \\ \hline 
Slope Parameter                         & $G_1 = 0.61 \pm 0.02$                                                                             & \cite{huang_photometric_2021}    \\
Slope Parameter                         & $G_2 = -0.006 \pm 0.006$                                                                            & \cite{huang_photometric_2021}    \\
Effective diameter                      & 1.32 $\pm 0.06$ km                                                                                  & \cite{ishiguro_polarimetric_2022}       \\
Absolute mag ($V$ filter)                            & $17.54\pm0.02$                                                                                             & \cite{ishiguro_polarimetric_2022}     \\
Sidereal period                         & $5.234246 \pm 0.000097$ hr                                                                             & This work  (Section \ref{sec:period})        \\
Bond albedo                             & 0.052                                                                                             & \cite{devogele_new_2020}      \\
Geometric Albedo                                  & 0.10 $\pm 0.02$                                                                                    & \cite{devogele_new_2020}      \\
Color $(g-r)$                             & $0.472\pm0.05$                                                                                              & This work (Section \ref{sec:heterogeneity}) \\
Color $(r-i)$                             & $0.065\pm0.05$                                                                                               & This work (Section \ref{sec:heterogeneity})    \\
Shape aspect ratio                      & 1.45                                                                                              & This work (Section \ref{sec:pole})     \\
Spin axis (preferred)                             & $\lambda_p = 116\fdegree6$, $\beta_p = -53\fdegree6$                                                                                         & This work (Section \ref{sec:pole})          \\
Spin axis                               & $\lambda = 300\fdegree3$, $\beta = -55\fdegree4$                                                                                                 & This work  (Section \ref{sec:pole})         \\
Critical density                        & 570 kg m$^{-3}$                                                                                               & \cite{jewitt_physical_2006}  \\
Photometric range                       & 0.4                                                                                               & \cite{jewitt_physical_2006}  \\  

Mass loss rate                              & $\lesssim 0.1$ kg m$^{-1}$                                                                                          & \cite{kasuga_wiseneowise_2022}    \\
\hline
 & Polarimetry$^{\dagger}$ &  \\ \hline 
Max degree of neg. polarization & $P_{min} = -1.2\% \pm 0.1\%$                                                                        & \cite{devogele_new_2020}      \\
Phase where $P_{min}$ occurs       & $\alpha_{min} = 9\fdegree5 \pm 0\fdegree2$                                                              & \cite{devogele_new_2020}      \\
Inversion angle                         & $\alpha_{inv} = 20\fdegree2 \pm 0\fdegree2$                                                     & \cite{devogele_new_2020}      \\
Slope at $\alpha_{inv}$                 & $h = 0.22\% \pm 0.01\%$                                                                             & \cite{devogele_new_2020}    \\
Geometric albedo                 & $p_{\mathrm{R}} = 0.1$                                                                             & \cite{ishiguro_polarimetric_2022}    \\
\end{tabular*}
\end{center}
\begin{center}
\textbf{Note. }$^{\dagger}$ Additional polarimetric results are contained in \cite{ishiguro_polarimetric_2022} that are too broad for this table.    
\end{center}
\end{table}

\clearpage


\section{\ud{} Observations}
\label{appendixB}

Tables \ref{Table:UDobs1}, \ref{Table:UDobs2}, and \ref{Table:UDobs3} describe all observations of \ud{} used in this work including those rejected from the shape modeling procedure.

\renewcommand\thefigure{\thesection\arabic{figure}}
\setcounter{figure}{0}
\renewcommand\thetable{\thesection.\arabic{table}}
\setcounter{table}{0}
\begin{deluxetable*}{lccccccccrhcc}
\tablecaption{Observational Circumstances}
\label{Table:UDobs1}
\tablehead{\colhead{\#} & \colhead{Date/Time}        & \colhead{Span}  & \colhead{Telescope}       & \colhead{Filter} & \colhead{$N_p$} & \colhead{$m_V$}    & \colhead{$r$}    & \colhead{$\Delta$} & \colhead{$\angle_{\mathrm{STO}}$} & \nocolhead{Observer}  & \colhead{Reference} & \nocolhead{End} \\ 
          & (UT)  & (h)  &      &       &       &   & (au)  & (au)  & ($^\circ$)  &       &      &   }
\tablecolumns{13}
\startdata
1 & 2005-Nov-02 11:54:32 & 3.26   & LOT         & $R$      & 6     & 18.3 & 1.37 & 0.55  & 36.7        & Kinoshita        & 1            \\
2 & 2005-Nov-03 12:11:24 & 3.22   & LOT          & $R$      & 5     & 18.4 & 1.39 & 0.57  & 36.7        & Kinoshita        & 1            \\
3 & 2005-Nov-04 12:05:20 & 3.00   & LOT          & $R$      & 13    & 18.5 & 1.4  & 0.59  & 36.7        & Kinoshita        & 1            \\
4 & 2005-Nov-05 11:52:45 & 1.78   & LOT          & $R$      & 16    & 18.6 & 1.41 & 0.61  & 36.6        & Kinoshita        & 1            \\
5& 2005-Nov-19 08:50:19  & 0.56   & UH88         & $R$      & 15    & 19.6 & 1.57 & 0.91  & 35.9        & Jewitt           & 2             \\
6 & 2005-Nov-20 07:23:39  & 0.79   & UH88          & $R$      & 16    & 19.7 & 1.58 & 0.93  & 35.9        & Jewitt           & 2             \\
7 & 2005-Nov-21 05:48:28  & 3.19   & UH88         & $R$      & 17    & 19.8 & 1.59 & 0.95  & 35.8        & Jewitt           & 2             \\
8 & 2005-Nov-22 05:31:34  & 3.53   & UH88         & $R$      & 9     & 19.9 & 1.61 & 0.99  & 35.7        & Jewitt           & 2            \\
9 & 2018-Sep-27 00:49:29  & 1.97   & LCO-fl16       & $r$      & 92    & 16.6 & 1.09 & 0.23  & 62.5        & ?                & 4            \\
10* & 2018-Sep-27 07:09:27  & 5.23   & 31in       & -      & 87    & 16.5 & 1.1  & 0.22  & 58.2        & ?                & 4            \\
11 & 2018-Sep-27 16:35:01 & 1.97   & LCO-fl11       & $r$      & 114   & 16.5 & 1.1  & 0.22  & 58.2        & ?                & 4            \\
12 & 2018-Oct-01 15:55:45 & 2.47   & LCO-fl11       & $r$      & 115   & 16.1 & 1.16 & 0.23  & 40.6        & ?                & 4            \\
13 & 2018-Oct-01 23:39:44 & 2.45   & LCO-fl16       & $r$      & 122   & 16.1 & 1.16 & 0.23  & 40.6        & ?                & 4            \\
14 & 2018-Oct-03 05:41:52  & 2.45   & LCO-fa15       & $r$      & 142   & 16   & 1.18 & 0.23  & 36.3        & ?                & 4            \\
15 & 2018-Oct-03 21:57:50 & 0.29   &     Ond\v{r}ejov     & $R$      & 17    & 15.9 & 1.2  & 0.24  & 28.0          & Pravec           & 5            \\
16 & 2018-Oct-03 23:22:49 & 1.46   &   Ond\v{r}ejov       & $R$      & 134   & 15.9 & 1.2  & 0.24  & 28.0          & Pravec           & 5            \\
17 & 2018-Oct-03 23:50:50 & 2.98   & LCO-fl16       & $r$      & 146   & 15.9 & 1.2  & 0.24  & 28.0          & ?                & 4            \\
18 & 2018-Oct-04 00:51:58  & 0.96   &     Ond\v{r}ejov     & $R$      & 86    & 15.9 & 1.2  & 0.24  & 28.0          & Pravec           & 5            \\
19 & 2018-Oct-04 02:20:05  & 1.63   &     Ond\v{r}ejov     & $R$      & 110   & 15.9 & 1.2  & 0.24  & 28.0          & Pravec           & 5            \\
20 & 2018-Oct-05 21:17:57 & 2.18   &     Ond\v{r}ejov    & $R$      & 158   & 15.9 & 1.22 & 0.24  & 24.1        & Pravec           & 5            \\
21 & 2018-Oct-05 23:32:08 & 1.97   &    Ond\v{r}ejov       & $R$      & 140   & 15.9 & 1.22 & 0.24  & 24.1        & Pravec           & 5            \\
22 & 2018-Oct-05 01:31:54  & 2.23   &  Ond\v{r}ejov        & $R$      & 160   & 15.9 & 1.22 & 0.24  & 24.1        & Pravec           & 5            \\
23* & 2018-Oct-06 04:07:48  & 8.38   & 31in       & $r$      & 32    & 15.9 & 1.23 & 0.25  & 20.4        & ?                & 4            \\
24 & 2018-Oct-06 02:59:19  & 6.29   & TRAPPIST-S & $R$      & 348   & 15.9 & 1.23 & 0.25  & 20.4        & ?                & 4            \\
25 & 2018-Oct-09 20:40:28 & 1.30   & Ond\v{r}ejov         & $R$      & 80    & 15.8 & 1.27 & 0.28  & 10.6        & Pravec           & 5            \\
26 & 2018-Oct-09 22:00:18 & 2.26   & Ond\v{r}ejov          & $R$      & 154   & 15.8 & 1.27 & 0.28  & 10.6        & Pravec           & 5     \\
27 & 2018-Oct-09 00:17:27  & 1.92   & Ond\v{r}ejov          & $R$      & 125   & 15.8 & 1.27 & 0.28  & 10.6        & Pravec           & 5            \\
28 & 2018-Oct-09 02:17:16  & 0.46   & Ond\v{r}ejov          & $R$      & 28    & 15.8 & 1.27 & 0.28  & 10.6        & Pravec           & 5            \\
29 & 2018-Oct-09 13:45:09 & 3.46   & LCO-fl11       & $r$      & 155   & 15.8 & 1.27 & 0.28  & 10.6        & ?                & 4            \\
30 & 2018-Oct-09 20:30:22 & 3.55   & LCO-fl06       & $r$      & 168   & 15.8 & 1.27 & 0.28  & 10.6        & ?                & 4            \\
\enddata
\raggedright
\footnotesize
\vspace{1mm}
\textbf{Notes.} Date/Time: start of observations; Span: duration of observations; $N_P$: number of points; $m_V$: \ud{} apparent $V-$band magnitude; $r$: Sun-target distance; $\Delta$: Earth---target distance; $\angle_{\mathrm{STO}}$: Sun--target--observer (phase) angle; LOT: Lulin One-meter Telescope; UH88: University of Hawaii 88-in Telescope; LCO: Las Cumbres Observatory; 31in: Lowell Observatory NURO 31-inch Telescope; Ond\v{r}ejov: Ond\v{r}ejov Observatory 0.65 m Telescope; TRAPPIST-S: South TRAnsiting Planets and PlanetesImals Small Telescope. \\
$^*$~Rejected from the inversion process due to excessive photometric noise or temporal overlap with other data.
\textbf{References.} (1) \cite{kinoshita_surface_2007}; (2) \cite{jewitt_physical_2006}; (3) \cite{warner_near-earth_2019}; (4) \cite{devogele_new_2020}; (5) this work.

\end{deluxetable*}

\setcounter{table}{1} 

\begin{deluxetable*}{lccccccccrhcc}
\tablecaption{Observational Circumstances}
\label{Table:UDobs2}
\tablehead{\colhead{\#} & \colhead{Date/Time}        & \colhead{Span}  & \colhead{Telescope}       & \colhead{Filter} & \colhead{$N_p$} & \colhead{$m_V$}    & \colhead{$r$}    & \colhead{$\Delta$} & \colhead{$\angle_{\mathrm{STO}}$} & \nocolhead{Observer}  & \colhead{Reference} & \nocolhead{End} \\ 
  & (UT)  & (h)  &      &       &       &     & (au)  & (au)  & ($^\circ$)  &      &      &   } 
\tablecolumns{12}
\startdata
31 & 2018-Oct-10 00:04:14  & 5.45   & TRAPPIST-N & $R$      & 285   & 15.8 & 1.29 & 0.29  & 7.7         & ?                & 4            \\
32 & 2018-Oct-10 00:30:22  & 2.69   & Ond\v{r}ejov          & $R$      & 154   & 15.8 & 1.29 & 0.29  & 7.7         & Pravec           & 5            \\
33* & 2018-Oct-10 03:09:27  & 9.02   & 31in       & -      & 57    & 15.8 & 1.29 & 0.29  & 7.7         & ?                & 4            \\
34 & 2018-Oct-10 04:59:36  & 2.28   & LCO-fa15       & $r$      & 112   & 15.8 & 1.29 & 0.29  & 7.7         & ?                & 4            \\
35 & 2018-Oct-12 20:38:08 & 3.17   & Ond\v{r}ejov          & $R$      & 107   & 15.8 & 1.31 & 0.31  & 2.6         & Pravec           & 5            \\
36 & 2018-Oct-12 23:49:46 & 2.09   & Ond\v{r}ejov          & $R$      & 114   & 15.8 & 1.31 & 0.31  & 2.6         & Pravec           & 5            \\
37 & 2018-Oct-12 06:00:36  & 1.34   & CS3         & $V$      & 26    & 15.8 & 1.31 & 0.31  & 2.6         & Stephens         & 3            \\
38 & 2018-Oct-12 08:15:57  & 2.54   & CS3          & $V$      & 26    & 15.8 & 1.31 & 0.31  & 2.6         & Stephens         & 3    \\
39 & 2018-Oct-13 05:10:00  & 1.90   & LCO-fa15       & $r$      & 88    & 15.7 & 1.33 & 0.33  & 0.6         & ?                & 4            \\
40 & 2018-Oct-13 12:04:43 & 3.36   & LCO-fl11       & $r$      & 153   & 15.7 & 1.33 & 0.33  & 0.6         & ?                & 4            \\
41 & 2018-Oct-14 19:55:12 & 0.46   & Ond\v{r}ejov          & $R$      & 19    & 15.9 & 1.34 & 0.34  & 2.1         & Pravec                & 5            \\
42 & 2018-Oct-14 12:13:49 & 2.50   & LCO-fl11       & $r$      & 122   & 15.9 & 1.34 & 0.34  & 2.1         & ?                & 4            \\
43 & 2018-Oct-15 18:47:03 & 2.69   & Ond\v{r}ejov          & $R$      & 120   & 16.2 & 1.35 & 0.35  & 4.1         & Pravec           & 5            \\
44 & 2018-Oct-15 21:29:53 & 2.54   & Ond\v{r}ejov          & $R$      & 121   & 16.2 & 1.35 & 0.35  & 4.1         & Pravec           & 5            \\
45* & 2018-Oct-15 02:26:25  & 9.10   & 31in       & $r$      & 48    & 16.2 & 1.35 & 0.35  & 4.1         & ?                & 4            \\
46 & 2018-Oct-15 03:32:17  & 1.49   & CS3          & $V$      & 27    & 16.2 & 1.35 & 0.35  & 4.1         & Stephens         & 3            \\
47 & 2018-Oct-15 05:07:32  & 1.68   & CS3         & $V$      & 37    & 16.2 & 1.35 & 0.35  & 4.1         & Stephens         & 3            \\
48 & 2018-Oct-15 06:56:53  & 1.80   & CS3          & $V$      & 31    & 16.2 & 1.35 & 0.35  & 4.1         & Stephens         & 3            \\
49 & 2018-Oct-15 09:02:17  & 1.90   & CS3          & $V$      & 37    & 16.2 & 1.35 & 0.35  & 4.1         & Stephens         & 3            \\
50 & 2018-Oct-15 19:22:50 & 1.75   & LCO-fl16       & $r$      & 60    & 16.2 & 1.35 & 0.35  & 4.1         & ?                & 4            \\
51 & 2018-Oct-16 01:59:55  & 3.98   & LCO-fa15       & $r$      & 171   & 16.4 & 1.36 & 0.37  & 5.9         & ?                & 4            \\
52* & 2018-Oct-16 02:24:27  & 3.17   & 31in       & $r$      & 14    & 16.4 & 1.36 & 0.37  & 5.9         & ?                & 4            \\
53 & 2018-Oct-16 03:13:13  & 1.75   & CS3          & $V$      & 22    & 16.4 & 1.36 & 0.37  & 5.9         & Stephens         & 3            \\
54 & 2018-Oct-16 05:00:16  & 1.54   & CS3          & $V$      & 27    & 16.4 & 1.36 & 0.37  & 5.9         & Stephens         & 3            \\
55 & 2018-Oct-16 06:42:44  & 3.26   & CS3          & $V$      & 52    & 16.4 & 1.36 & 0.37  & 5.9         & Stephens         & 3            \\
56 & 2018-Oct-16 09:59:22  & 0.74   & CS3          & $V$      & 15    & 16.4 & 1.36 & 0.37  & 5.9         & Stephens         & 3            \\
57 & 2018-Oct-16 12:15:49 & 0.36   & LCO-fl12       & $r$      & 14    & 16.4 & 1.36 & 0.37  & 5.9         & ?                & 4            \\
58 & 2018-Oct-16 19:59:50 & 3.48   & LCO-fl16       & $r$      & 132   & 16.4 & 1.36 & 0.37  & 5.9         & ?                & 4            \\
59 & 2018-Oct-16 21:42:10 & 2.71   & Ond\v{r}ejov          & $R$      & 23    & 16.6 & 1.38 & 0.38  & 7.6         & Pravec           & 5            \\
60 & 2018-Oct-17 02:59:36  & 3.46   & LCO-fa15       & $r$      & 150   & 16.6 & 1.38 & 0.38  & 7.6         & ?                & 4            \\
\enddata
\raggedright
\footnotesize
\vspace{1mm}
\textbf{Notes.} Date/Time: start of observations; Span: duration of observations; $N_P$: number of points; $m_V$: \ud{} apparent $V-$band magnitude; $r$: Sun-target distance; $\Delta$: Earth---target distance; $\angle_{\mathrm{STO}}$: Sun--target--observer (phase) angle; TRAPPIST-N: North TRAnsiting Planets and PlanetesImals Small Telescope; Ond\v{r}ejov: Ond\v{r}ejov Observatory 0.65 m Telescope; Lowell Observatory NURO 31-inch Telescope; LCO: Las Cumbres Observatory; CS3: Center for Solar System Studies. \\
$^*$~Rejected from the inversion process due to excessive photometric noise or temporal overlap with other data. \\
\textbf{References.} (1) \cite{kinoshita_surface_2007}; (2) \cite{jewitt_physical_2006}; (3) \cite{warner_near-earth_2019}; (4) \cite{devogele_new_2020}; (5) this work.

\end{deluxetable*}

\begin{deluxetable*}{lccccccccrhcc}
\tablecaption{Observational Circumstances}
\label{Table:UDobs3}
\tablehead{\colhead{\#} & \colhead{Date/Time}        & \colhead{Span}  & \colhead{Telescope}       & \colhead{Filter} & \colhead{$N_p$} & \colhead{$m_V$}    & \colhead{$r$}    & \colhead{$\Delta$} & \colhead{$\angle_{\mathrm{STO}}$} & \nocolhead{Observer}  & \colhead{Reference} & \nocolhead{End} \\ 
  & (UT)  & (h)  &      &       &       &     & (au)  & (au)  & ($^\circ$)  &      &      &   } 
\tablecolumns{12}
\startdata
61 & 2018-Oct-17 04:20:39  & 3.12   & TRAPPIST-N & $R$      & 179   & 16.6 & 1.38 & 0.38  & 7.6         & ?                & 4            \\
62 & 2018-Oct-17 18:52:22 & 2.30   & Ond\v{r}ejov          & $R$      & 76    & 16.7 & 1.39 & 0.4   & 9.3         & Pravec           & 5           \\
63 & 2018-Oct-17 21:11:36 & 2.69   & Ond\v{r}ejov          & $R$      & 91    & 16.7 & 1.39 & 0.4   & 9.3         & Pravec           & 5            \\
64 & 2018-Oct-20 23:58:39 & 2.28   & TRAPPIST-N & $r$      & 133   & 17   & 1.41 & 0.43  & 12.1        & ?                & 4            \\
65 & 2018-Oct-20 01:41:08  & 2.30   & LCO-fa15       & $r$      & 91    & 17   & 1.41 & 0.43  & 12.1        & ?                & 4            \\
66 & 2018-Oct-25 10:28:53 & 3.77   & LCO-fl12       & $r$      & 75    & 17.7 & 1.47 & 0.51  & 17.8        & ?                & 4            \\
67 &2018-Oct-27 01:43:35  & 2.98   & LCO-fa15       & $r$      & 69    & 17.9 & 1.5  & 0.55  & 19.6        & ?                & 4            \\
68 & 2018-Oct-28 00:08:31  & 3.43   & LCO-fa03       & $r$      & 84    & 18.1 & 1.5  & 0.56  & 20.4        & ?                & 4            \\
69 & 2018-Oct-30 10:38:14 & 3.46   & LCO-fl11       & $r$      & 73    & 18.3 & 1.53 & 0.6   & 21.9        & ?                & 4            \\
70 & 2018-Oct-31 10:24:58 & 3.02   & LCO-fl11       & $r$      & 60    & 18.4 & 1.54 & 0.62  & 22.6        & ?                & 4            \\
71 & 2018-Nov-01 01:42:22  & 2.93   & LCO-fa15       & $r$      & 62    & 18.5 & 1.55 & 0.64  & 23.2        & ?                & 4            \\
72 & 2018-Nov-02 02:43:58  & 1.90   & LCO-fa15       & $r$      & 43    & 18.6 & 1.56 & 0.67  & 23.8        & ?                & 4            \\
73 & 2018-Nov-04 00:24:18  & 4.06   & Danish          & $R$      & 20    & 18.8 & 1.58 & 0.69  & 24.8        & Pravec           & 5            \\
$74^+$ & 2019-Oct-18 09:19:18  & 3.24   & LDT        & $VR$     & 83    & 20.3 & 1.66 & 1.16  & 36.3        & Devogele         & 5            \\
75 & 2019-Nov-03 01:20:13  & 5.16   & NOT        & $r$      & 329   & 19.6 & 1.49 & 0.87  & 39.6        & Granvik          & 5            \\
76 & 2019-Nov-18 07:24:19  & 5.81   & LDT        & $VR$     & 666   & 18.8 & 1.32 & 0.66  & 45.2       & Moskovitz, Kueny & 5            \\
77 & 2019-Nov-18 01:15:07  & 5.47   & NOT        & $gri$      & 254   & 18.8 & 1.33 & 0.66  & 45.8       & Granvik          & 5            \\
78 & 2019-Nov-23 00:25:28  & 5.52   & TRAPPIST-N & $Rc$     & 130   & 18.6 & 1.25 & 0.58  & 50.7       & Jehin, Ferrais   & 5            \\
79 & 2019-Nov-24 02:10:49  & 2.57   & TRAPPIST-N & $Rc$     & 62    & 18.5 & 1.23 & 0.56  & 51.7       & Jehin, Ferrais   & 5            \\
80* & 2019-Nov-25 00:34:05  & 5.57   & TRAPPIST-N & $Rc$     & 115    & 18.5 & 1.22 & 0.55  & 52.7       & Jehin, Ferrais   & 5            \\
81 & 2021-Oct-27 03:30:14  & 0.65   & Danish & $R$     & 8   & 19.2 & 1.90 & 0.92  & 7.2       & Pravec           &  5 \\
82 & 2021-Oct-28 03:11:31  & 1.27   & Danish & $R$     & 13   & 19.3 & 1.91 & 0.93  & 7.7       & Pravec           &  5 \\
83 & 2021-Oct-29 00:05:46  & 3.58   & Danish & $R$     & 19   & 19.3 & 1.91 & 0.94  & 8.2       & Pravec           &  5 \\
84 & 2021-Oct-30 00:02:53  & 3.56   & Danish & $R$     & 25   & 19.4 & 1.92 & 0.95  & 8.7       & Pravec           &  5 \\
85 & 2021-Nov-03 01:29:31  & 2.47   & LDT & $VR$     & 115   & 19.6 & 1.95 & 1.00  & 11.2       & Kareta          &  5 \\
86 & 2005---2011 SPARSE            &   ---    &      CSS      &     $V$   &    19   &   ---   &    ---  &   ---    &     ---        &          ---        &  6 \\
87 & 2014---2021 SPARSE            &   ---    &      PS      &     $w$   &    31   &   ---   &    ---  &   ---    &     ---        &          ---        &  7 \\
88 & 2018---2021 SPARSE            &   ---    &      ZTF      &     $V$   &    71   &   ---   &    ---  &   ---    &     ---        &          ---        &  8 \\
89 & 2005---2011 SPARSE            &   ---    &      ATLAS      &     $o$   &    41   &   ---   &    ---  &   ---    &     ---        &          ---        &  9 \\
90 & 2005---2011 SPARSE            &   ---    &      CSS      &     $G$   &    44   &   ---   &    ---  &   ---    &     ---        &          ---        &  6 \\
\enddata
\raggedright
\footnotesize
\vspace{1mm}
\textbf{Notes.} Date/Time: start of observations; Span: duration of observations; $N_P$: number of points; $m_V$: \ud{} apparent $V-$band magnitude; $r$: Sun-target distance; $\Delta$: Earth---target distance; $\angle_{\mathrm{STO}}$: Sun--target--observer (phase) angle; TRAPPIST-N: North TRAnsiting Planets and PlanetesImals Small Telescope; Ond\v{r}ejov: Ond\v{r}ejov Observatory 0.65 m Telescope; LCO: Las Cumbres Observatory; LDT: Lowell Discovery Telescope; NOT: Nordic Optical Telescope; Danish: La Silla Observatory 1.54 m Telescope; CSS: Catalina Sky Survey; PS: Pan-STARRS; ZTF: Zwicky Transient Facility; ATLAS: Asteroid Terrestrial-impact Last Alert System. \\
$^*$~Rejected from the inversion process due to excessive photometric noise or temporal overlap with other data. \\
$^+$ First 1.5 hr rejected from the inversion process due to being inconsistent with our best-fit models. \\
\textbf{References.} (1) \cite{kinoshita_surface_2007}; (2) \cite{jewitt_physical_2006}; (3) \cite{warner_near-earth_2019}; (4) \cite{devogele_new_2020}; (5) This work; (6) \cite{larson_css_2003}; (7) \citep{chambers_pan-starrs1_2016}; (8) \cite{bellm_zwicky_2019}; (9) \cite{tonry_atlas_2018}.

\end{deluxetable*}

\end{document}